\documentclass[ALICE,manyauthors]{cernphprep}

\usepackage{graphicx}
\usepackage{textcomp}
\usepackage[utf8]{inputenc}
\usepackage[T1]{fontenc}
\usepackage{gensymb}

\usepackage{epsfig}
\usepackage{hyperref}
\usepackage{url}
\usepackage{multirow}
\usepackage{rotating}
\usepackage{epstopdf}

\usepackage{lineno}

\newcommand{\jpsi}{{{\rm J}/\psi}}
\newcommand{\psiprime}{{\psi({\rm 2S})}}
\newcommand{\chic}{{\chi_c}}
\newcommand{\chib}{{\chi_b}}
\newcommand{\pp}{{\rm pp}}
\newcommand{\ppbar}{{{\rm p}\overline{\rm p}}}
\newcommand{\pt}{{p_{\rm T}}}
\newcommand{\aeff}{{A\epsilon}}
\newcommand{\aeffmean}{{\langle\aeff\rangle}}
\newcommand{\mmumu}{{m_{\mu\mu}}}
\RequirePackage{lineno}
\begin{document}

%
%
\begin{titlepage}
\PHnumber{2014-042}                
\PHdate{March 5, 2014}             
%
%
\title{Measurement of quarkonium production at forward rapidity in $\mathbf{pp}$ collisions at $\mathbf{\sqrt{s}=7}~$TeV}
\ShortTitle{Quarkonium production at $\mathbf{\sqrt{s}=7}~$TeV}   
%
\Collaboration{ALICE Collaboration%
         \thanks{See Appendix~\ref{app:collab} for the list of collaboration
                      members}}
\ShortAuthor{ALICE Collaboration}      
\begin{abstract}
The inclusive production cross sections at forward rapidity of $\jpsi$, $\psiprime$, $\rm\Upsilon$(1S) and $\rm\Upsilon$(2S) are measured in $\pp$ collisions at $\sqrt{s}=7~{\rm TeV}$ with the ALICE detector at the LHC. The analysis is based on a data 
sample corresponding to an integrated luminosity of 1.35 pb$^{-1}$. Quarkonia are reconstructed
in the dimuon-decay channel and the signal yields are evaluated by fitting the $\mu^+\mu^-$ invariant mass distributions.
The differential production cross sections are measured as a function of the transverse momentum $\pt$ and rapidity $y$, over the ranges $0<\pt<20$~GeV/$c$ for $\jpsi$, $0<\pt<12$~GeV/$c$ for all other resonances and for $2.5<y<4$. The measured cross sections integrated over $\pt$ and $y$, and assuming unpolarized quarkonia, are: $\sigma_{\rm \jpsi}=6.69\pm0.04\pm0.63$~$\mu$b, $\sigma_\psiprime=1.13\pm0.07\pm0.19$~$\mu$b, $\sigma_{\rm\Upsilon({\rm 1S})}=54.2\pm5.0\pm6.7$~nb and $\sigma_{\rm\Upsilon({\rm 2S})}=18.4\pm3.7\pm2.9$~nb, where the first uncertainty is statistical and the second one is systematic. The results are compared to measurements performed by other LHC experiments and to theoretical models.

\end{abstract}
\end{titlepage}
\setcounter{page}{2}
%
\section{Introduction}
\label{intro}
Quarkonia are bound states of either a charm and anti-charm quark pair (charmonia, e.g. $\jpsi$, $\chi_c$ and $\psiprime$) or a bottom and anti-bottom quark pair (bottomonia, e.g. $\rm\Upsilon$(1S), $\rm\Upsilon$(2S), $\chi_b$ and $\rm\Upsilon$(3S)). While the production of the heavy quark pairs in $\pp$ collisions is relatively well understood in the context of perturbative QCD calculations~\cite{Cacciari:1998it,Cacciari:2001td,Cacciari:2012ny}, their binding into quarkonium states is inherently a non-perturbative process and the understanding of their production in hadronic collisions remains unsatisfactory despite the availability of large amounts of data and the considerable theoretical progress made in recent years~\cite{Brambilla:2010cs}.  For instance none of the models are able to describe simultaneously different aspects of quarkonium production such as polarization, transverse momentum and energy dependence of the cross sections.

There are mainly three approaches used to describe the hadronic production of quarkonium: the Color-Singlet Model (CSM), the Color Evaporation Model (CEM) and the Non-Relativistic QCD (NRQCD) framework. 

In the CSM~\cite{Einhorn:1975ua,Carlson:1976cd,Baier:1981uk}, perturbative QCD is used to model the production of on-shell heavy quark pairs, with the same quantum numbers as the quarkonium into which they hadronize. This implies that only color-singlet quark pairs are considered. Historically, CSM calculations performed at leading order (LO) in $\alpha_s$, the strong interaction coupling constant, have been unable to reproduce the magnitude and the $\pt$ dependence of the $\jpsi$ production cross section measured by CDF at the Tevatron~\cite{Abe:1997jz}. Several improvements to the model have been worked out since then: the addition of all next-to-leading order (NLO) diagrams~\cite{PhysRevLett.98.252002} as well as some of the next-to-next-to-leading order (NNLO)~\cite{Artoisenet:2008fc,Lansberg:2008gk}; the inclusion of other processes to the production of high $\pt$ quarkonia such as gluon fragmentation~\cite{Braaten:1994xb} or the production of a quarkonium in association with a heavy quark pair~\cite{Artoisenet:2007xi} and the relaxation of the requirement that the heavy quark pair is produced on-shell before hadronizing into the quarkonium~\cite{Haberzettl:2007kj}. All these improvements contribute to a better agreement between theory and data but lead to considerably larger theoretical uncertainties and/or to the introduction of extra parameters that are fitted to the data.

In the CEM~\cite{Halzen:1977rs,Fritzsch:1977ay,Amundson:1996qr}, the production cross section of a given quarkonium state is considered proportional to the cross section of its constituting heavy quark pair, integrated from the sum of the masses of the two heavy quarks to the sum of the masses of the lightest corresponding mesons (D or B). The proportionality factor for a given quarkonium state is assumed to be universal and independent of its transverse momentum $\pt$ and rapidity $y$. It follows that the ratio between the yields of two quarkonium states formed out of the same heavy quarks is independent of the collision energy as well as of $\pt$ and $y$. This model is mentioned here for completeness but is not confronted to the data presented in this paper.

Finally, in the framework of NRQCD~\cite{Bodwin:1994jh}, contributions to the quarkonium cross section from the heavy-quark pairs produced in a color-octet state are also taken into account, in addition to the color-singlet contributions described above. The neutralization of the color-octet state into a color-singlet is treated as a non-perturbative process. It is expanded in powers of the relative velocity between the two heavy quarks and parametrized using universal long-range matrix elements which are considered as free parameters of the model and fitted to the data. This approach has recently been extended to NLO~\cite{Ma:2010jj,Butenschoen:2010rq,Xu:2012am} and is able to describe consistently the production cross section of quarkonia in $\ppbar$ and $\pp$ collisions at Tevatron, RHIC and, more recently, at the LHC. However, NRQCD predicts a sizable transverse component to the polarization of the $\jpsi$ meson, which is in contradiction with the data measured for instance at Tevatron~\cite{Abulencia:2007us} and at the LHC~\cite{alice:2012pol,Aaij:2013nlm,Chatrchyan:2013cla,Chatrchyan:2012woa}.

Most of the observations and discrepancies described above apply primarily to charmonium production. For bottomonium production, theoretical calculations are more robust due to the higher mass of the bottom quark and the disagreement between data and theory is less pronounced than in the case of charmonium~\cite{Abe:1995an,LHCb:2012aa}. Still, the question of a complete and consistent description of the production of all quarkonium states remains open and the addition of new measurements in this domain will help constraining the various models at hand.

In this paper we present measurements of the inclusive production cross section of several quarkonium states (namely $\jpsi$, $\psiprime$, $\rm\Upsilon$(1S) and $\rm\Upsilon$(2S)) using the ALICE detector at forward rapidity ($2.5<y<4$) in  $\pp$ collisions at $\sqrt{s}=7$~TeV. Inclusive measurements contain, in addition to the quarkonium direct production, contributions from the decay of higher mass excited states: predominantly $\psiprime$ and $\chic$ for the $\jpsi$; $\rm\Upsilon$(2S), $\chib$ and $\rm\Upsilon$(3S) for the $\rm\Upsilon$(1S), and $\rm\Upsilon$(3S) and $\chib$ for the $\rm\Upsilon$(2S). For $\jpsi$ and $\psiprime$, they contain as well contributions from non-prompt production, mainly from the decay of $b$-mesons. For the $\jpsi$ meson, these measurements represent an increase by a factor of about 80 in terms of luminosity with respect to published ALICE results~\cite{Aamodt:2011gj,Aamodt:2011gjE}. For the $\psiprime$ and the $\rm\Upsilon$, we present here the first ALICE measurements in $\pp$ collisions.

This paper is organized as follows: a brief description of the ALICE detectors used for this analysis and of the data sample is provided in Section~\ref{sec:apparatus}; the analysis procedure is described in Section~\ref{sec:analysis}; in Section~\ref{sec:results} the results are presented and compared to those obtained by other LHC experiments; finally, in Section~\ref{sec:theory} the results are compared to several theoretical calculations.


\section{\label{sec:apparatus}Experimental apparatus and data sample}

\subsection{Experimental apparatus}
The ALICE detector is extensively described in~\cite{ALICE}. The analysis presented in this paper is based on muons detected at forward pseudo-rapidity ($-4<\eta<-2.5$) in the muon spectrometer~\cite{Aamodt:2011gj}\footnote{In the ALICE reference frame the muon spectrometer covers negative $\eta$. However, we use positive values when referring to $y$.}.
In addition to the muon spectrometer, the Silicon Pixel Detector (SPD)~\cite{ALICE_ITS} and the V0 scintillator hodoscopes~\cite{Abbas:2013taa} are used to provide primary vertex reconstruction and a Minimum Bias (MB) trigger, respectively. The T0 \v{C}erenkov detectors~\cite{Bondila:2005xy} are also used for triggering purposes and to evaluate some of the systematic uncertainties on the integrated luminosity determination. The main features of these detectors are listed in the following paragraphs.

The muon spectrometer consists of a front absorber followed by a 3~Tm dipole magnet, coupled to tracking and triggering devices. The front absorber, made of carbon, concrete and steel is placed between 0.9 and 5 m from the Interaction Point (IP). It filters muons from hadrons, thus decreasing the occupancy in the first stations of the tracking system. Muon tracking is performed by means of five stations, positioned between 5.2 and 14.4~m from the IP, each one consisting of two planes of Cathode Pad Chambers. The total number of electronic channels is close to 1.1$\times10^{6}$ and the intrinsic spatial resolution is about 70~$\mu$m in the bending direction. The first and the second stations are located upstream of the dipole magnet, the third station is embedded inside its gap and the fourth and the fifth stations are placed downstream of the dipole, just before a 1.2 m thick iron wall (7.2 interaction lengths) which absorbs hadrons escaping the front absorber and low momentum muons (having a total momentum $p<1.5$~GeV/$c$ at the exit of the front absorber). The muon trigger system is located downstream of the iron wall and consists of two stations positioned at 16.1 and 17.1 m from the IP, each equipped with two planes of Resistive Plate Chambers (RPC). The spatial resolution achieved by the trigger chambers is better than 1~cm, the time resolution is about 2~ns and the efficiency is higher than 95\%~\cite{mtrBossu2012}. The muon trigger system is able to deliver single and dimuon triggers above a programmable $\pt$ threshold, via an algorithm based on the RPC spatial information~\cite{mtr2006}. For a given trigger configuration, the threshold is defined as the $\pt$ value for which the single muon trigger efficiency reaches 50\%~\cite{mtrBossu2012}. Throughout its entire length, a conical absorber ($\theta<2\degree$) made of tungsten, lead and steel, shields the muon spectrometer against secondary particles produced by the interaction of large-$\eta$ primary particles in the beam pipe. 

Primary vertex reconstruction is performed using the SPD, the two innermost layers of the Inner Tracking System (ITS)~\cite{ALICE_ITS}. It covers the pseudo-rapidity ranges $|\eta|<2$ and $|\eta|<1.4$, for the inner and outer layers respectively. The SPD has in total about $10^7$ sensitive pixels on 240 silicon ladders, aligned using $\pp$ collision data as well as cosmic rays to a precision of 8 $\mu$m.

The two V0 hodoscopes, with 32 scintillator tiles each, are placed on opposite sides of the IP, covering the pseudo-rapidity ranges $2.8<\eta< 5.1$ and $-3.7<\eta<-1.7$. Each hodoscope is segmented into eight sectors and four rings of equal azimuthal and pseudo-rapidity coverage, respectively. A logical AND of the signals from the two hodoscopes constitutes the MB trigger, whereas the timing information of the two is used offline to reject beam-halo and beam-gas events, thanks to the intrinsic time resolution of each hodoscope which is better than 0.5~ns.

The T0 detectors are two arrays of 12 quartz \v{C}erenkov counters, read by photomultiplier tubes and located on opposite sides of the IP, covering the pseudo-rapidity ranges $4.61<\eta<4.92$ and $-3.28<\eta<-2.97$, respectively. They measure the time of the collision with a precision of $\sim 40$~ps in $\pp$ collisions and this information can also be used for trigger purposes.

\subsection{\label{subsec:luminosity}Data sample and integrated luminosity}

The data used for the analysis were collected in 2011. About 1300 proton bunches were circulating in each LHC ring and the number of bunches colliding at the ALICE IP
was ranging from 33 to 37. The luminosity was adjusted by means of the beam separation in the transverse (horizontal) direction to a value of $\sim 2\times10^{30}$~cm$^{-2}$s$^{-1}$. The average number of interactions per bunch crossing in such conditions is about~0.25, corresponding to a pile-up probability of $\sim$12\%. 
The trigger condition used for data taking is a dimuon-MB trigger formed by the logical AND of the MB trigger and an unlike-sign dimuon trigger with a $\pt$ threshold of 1~GeV/$c$ for each of the two muons. 

About 4$\times$10$^{6}$~dimuon-MB-triggered events were analyzed, corresponding to an integrated luminosity $L_{\rm{int}}=1.35\pm0.07$~pb$^{-1}$. The integrated luminosity is calculated on a run-by-run basis using the MB trigger counts measured with scalers before any data acquisition veto, divided by the MB trigger cross section and multiplied by the dimuon-MB trigger lifetime (75.6\% on average). The MB trigger counts are corrected for the trigger purity (fraction of events for which the V0 signal arrival times on the two sides lie in the time window corresponding to beam-beam collisions) and for pile-up. The MB trigger cross section is measured with the van der Meer (vdM) scan method~\cite{vanderMeer:1968zz}. The result of the vdM scan measurement~\cite{Abelev:2012sea} is corrected by a factor $0.990\pm0.002$ arising from a small modification of the V0 high voltage settings which occurred between the vdM scan and the period when the data were collected. The resulting trigger cross section is $\sigma_{\rm MB}=53.7\pm1.9({\rm syst})$~mb.

\section{\label{sec:analysis}Data analysis}

The quarkonium production cross section $\sigma$ is determined from the number of reconstructed quarkonia $N$ corrected by the branching ratio in dimuon ${\rm BR}_{\mu^+\mu^-}$ and the mean acceptance times efficiency $\aeffmean$ to account for detector effects and analysis cuts. 
The result is normalized to the integrated luminosity $L_{\rm{int}}$:
\begin{eqnarray}
\sigma=\frac{1}{L_{\rm{int}}} \frac{N}{{\rm BR}_{\mu^+\mu^-} \times \aeffmean}, 
\end{eqnarray}
with ${\rm BR}_{\mu^+\mu^-}=(5.93\pm0.06)$\%, $(0.78\pm0.09)$\%, $(2.48\pm0.05)$\% and $(1.93\pm0.17)$\% for $\jpsi$, $\psiprime$, $\rm\Upsilon$(1S) and $\rm\Upsilon$(2S), respectively~\cite{PhysRevD.86.010001}. Pile up events have no impact on the reconstruction of the quarkonium yields and are properly accounted for by the luminosity measurement.

\subsection{\label{ssec:signal}Signal extraction}

Quarkonia are reconstructed in the dimuon decay channel and the signal yields are evaluated using a fit to the $\mu^+\mu^-$ invariant mass distributions, as detailed in~\cite{Aamodt:2011gj}. In order to improve the purity of the dimuon sample, the following selection criteria are applied: 
\begin{itemize}
\item both muon tracks in the tracking chambers must match a track reconstructed in the trigger system; 
\item tracks are selected in the pseudo-rapidity range $-4\leq\eta\leq-2.5$;
\item the transverse radius of the track, at the end of the front absorber, is in the range $17.6\leq R_{\rm{abs}} \leq 89.5$~cm;
\item the dimuon rapidity is in the range 2.5 $\leq$ y $\leq$ 4;
\item a cut on the product of the total momentum of a given track and its distance to the primary vertex in the tranverse plane (called DCA) is applied for the bottomonium analysis in order to reduce the background under the $\rm\Upsilon$ signals. It is set to $6\times\sigma_{\rm pDCA}$, where $\sigma_{\rm pDCA}$ is the resolution on this quantity.
The cut accounts for the total momentum and angular resolutions of the muon detector as well as for the multiple scattering in the front absorber. 
This cut is not applied to the $\jpsi$ and $\psiprime$ analyses because it has negligible impact on the signal-to-background ratio for these particles.
\end{itemize}
These selection criteria help in removing hadrons escaping from (or produced in) the front absorber, low-$\pt$ muons from pion and kaon decays, secondary muons produced in the front absorber and fake muon tracks,  without significantly affecting the signals. Applying this selection criteria improves the signal-to-background ratio by 30\% for the $\jpsi$ and by a factor two for the $\psiprime$. It also allows to reduce the background by a factor three in the $\rm\Upsilon$ mass region.
 
The $\jpsi$ and $\psiprime$ yields are evaluated by fitting the dimuon invariant mass distribution in the mass range $2 < \mmumu < 5$~GeV/$c^2$. The function used in the fit is the sum of either two extended Crystal Ball (CB2) functions\footnote{The Crystal Ball function consists of a Gaussian core and a power law tail at low masses, as defined in~\cite{Gaiser:1982yw}. The CB2 function extends the standard Crystal Ball function by a second power law tail for high masses.}~\cite{Gaiser:1982yw} or two pseudo-Gaussian functions \cite{Shahoyan:2001sd} for the signals. The background is described by either a Gaussian with a width that varies linearly with the mass, also called Variable Width Gaussian (VWG), or the product of a fourth order polynomial function and an exponential function (Pol4 $\times$ Exp). 

\begin{figure}[!hbt]
\includegraphics[width=0.5\textwidth]{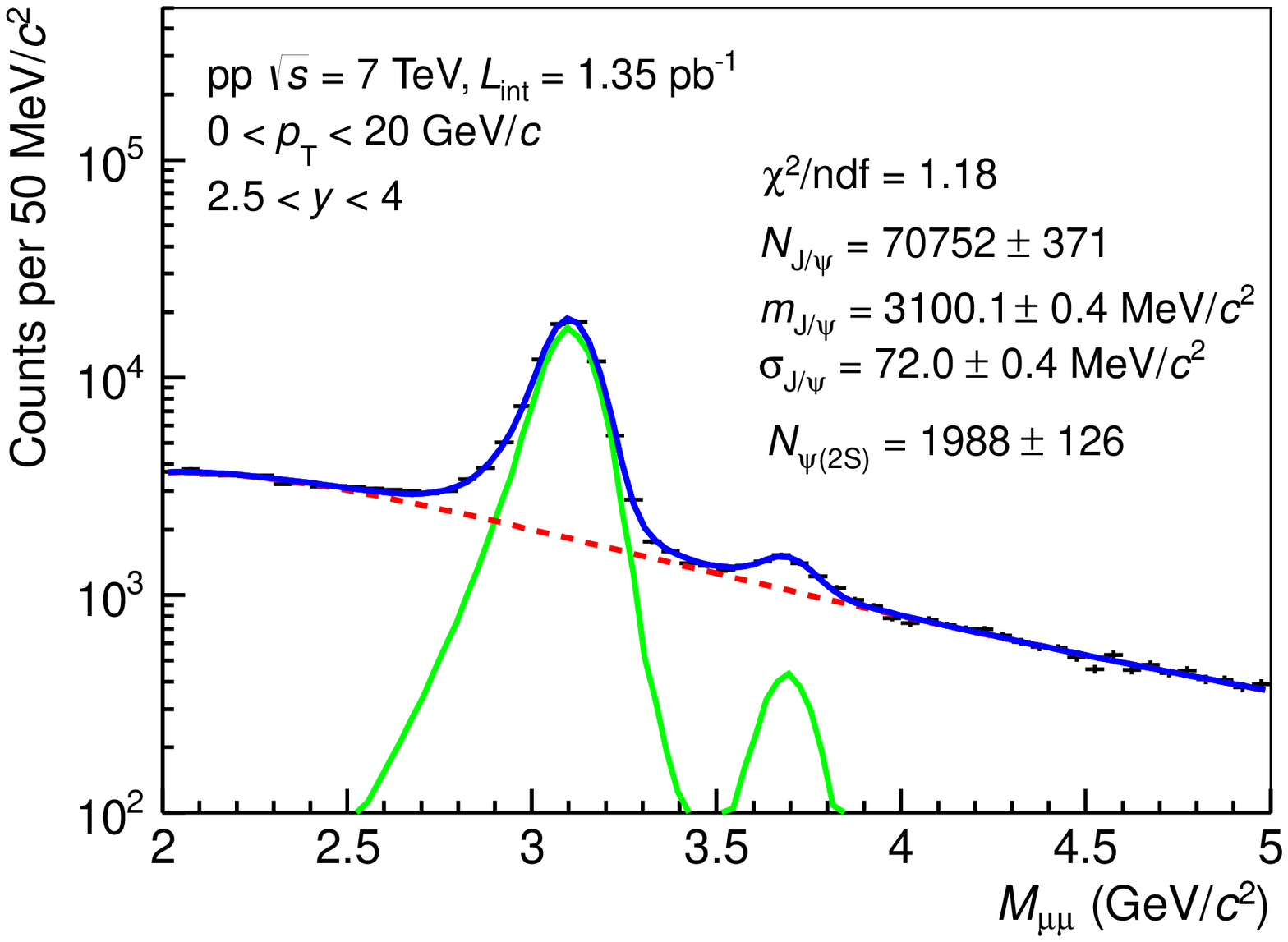}
\includegraphics[width=0.5\textwidth]{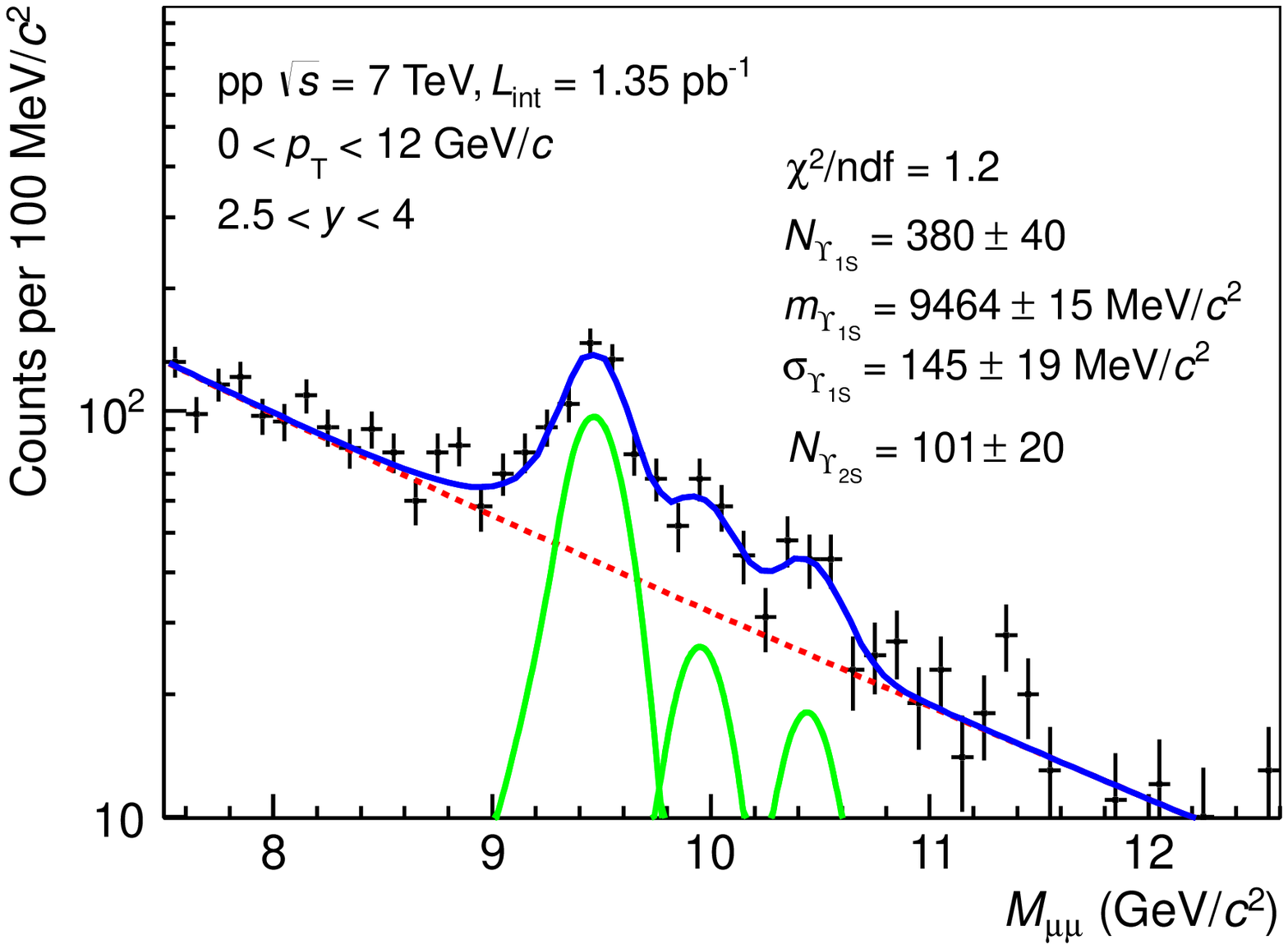}
\caption{Dimuon invariant mass distribution in the region of charmonia (left) and bottomonia (right). Solid (dotted) lines correspond to signal (background) fit functions. The sum of the various fit functions is also shown as a solid line. For the $\jpsi$ and $\psiprime$, a combination of two extended Crystal Ball functions is used for the signal and a variable width Gaussian function is used for the background. For the $\rm\Upsilon$ resonances, a combination of extended Crystal Ball functions is used for the signals and two power law functions for the background.}
\label{imass}
\end{figure}

The normalization factors of the signal functions are left free, together with the position and the width of the $\jpsi$ signal. On the other hand, the position and the width of the $\psiprime$ are tied to the corresponding parameters of the $\jpsi$ by forcing the mass difference between the two states to be equal to the one given by the Particle Data Group~\cite{PhysRevD.86.010001} and the mass resolution ratio to match the value obtained from a Monte Carlo (MC) simulation. The tail parameters for the $\jpsi$ are determined by fitting the shape of the $\jpsi$ signal obtained from the simulation. The same tail parameters are used for the $\psiprime$ as the resonances are separated by only 590~MeV/$c^{2}$ so that the energy straggling and multiple Coulomb scattering effects of the front absorber on the decay muons are expected to be similar. All the parameters of the functions used to fit the background are left free. An example of fit to the dimuon invariant mass distribution in the $\jpsi$ and $\psiprime$ mass region is shown in the left panel of Fig.~\ref{imass}.

The $\rm\Upsilon$(1S), (2S) and (3S) signal extractions are performed as for the $\jpsi$ and $\psiprime$ by fitting the dimuon invariant mass distribution in the mass range $5 < \mmumu < 15$~GeV/$c^2$. Due to the limited statistics, only the $\rm\Upsilon$(1S) and $\rm\Upsilon$(2S) yields are measured in this analysis. The background is fitted with a sum of either two power law or two exponential functions with all parameters left free. Each of the three $\rm\Upsilon$ signals (1S, 2S and 3S) is fitted with a Gaussian or a CB2 function. The fit parameters of the $\rm\Upsilon$(1S) signal are left free, whereas the width and mass position of the $\rm\Upsilon$(2S) and $\rm\Upsilon$(3S) are fixed with respect to the ones of the $\rm\Upsilon$(1S) in the same way as the $\psiprime$ parameters are fixed to the $\jpsi$. For the CB2 fit, the tail parameters of the function are fixed using the same method as for the charmonium signal extraction. An example of fit to the dimuon invariant mass distribution in the $\rm\Upsilon$'s mass region is shown in the right panel of Fig.~\ref{imass}. 

About 70800 $\jpsi$, 2000 $\psiprime$, 380 $\rm\Upsilon$(1S) and 100 $\rm\Upsilon$(2S) have been measured with signal-to-background ratios (S/B), evaluated within three standard deviations with respect to the quarkonium pole mass, of 4, 0.2, 1 and 0.3, respectively. 

In order to determine the $\pt$ differential cross sections, the data sample is divided in thirteen, nine and five transverse momentum intervals for $\jpsi$, $\psiprime$ and $\rm\Upsilon$(1S), respectively. The differential cross section as a function of rapidity is evaluated in six intervals for the $\jpsi$ and $\psiprime$ and three for the $\rm\Upsilon$(1S). Given the available statistics, only the measurement of the $\pt$- and $y$-integrated $\rm\Upsilon$(2S) cross section is possible. 
The quarkonium raw yields obtained from the differential study are reported in Section~\ref{sec:tables}. 
For $\jpsi$, the S/B ratio increases from $2.2$ to $8.5$ with increasing $\pt$ and from $3.7$ to $5.4$ with increasing rapidity. For $\psiprime$, it increases from $0.1$ to $0.6$ with increasing $\pt$ and from $0.1$ to $0.2$ with increasing rapidity. For the $\rm\Upsilon$(1S), no variation of the S/B ratio is observed within statistical uncertainties.

\subsection{Acceptance and efficiency corrections}

The measured yields obtained from the fits to the dimuon invariant mass distributions are corrected by the acceptance times efficiency factor $\aeffmean$ to determine the production yields of the four resonances.

In order to evaluate the $\aeffmean$ factor, simulations of quarkonium production in $\pp$ collisions at $\sqrt{s}=7~{\rm TeV}$ are performed with realistic $\pt$ and $y$ distributions, obtained by fitting existing data measured at the same energy for $\jpsi$ and $\psiprime$~\cite{Aaij:2011jh,Aaij:2012ag}, and by scaling CDF data~\cite{Abe:1995an} to $\sqrt{s}=7$~TeV for the $\rm\Upsilon$. All resonances are forced to decay into two muons. Particle transport is performed using GEANT3~\cite{GEANT3} and a realistic detector response is applied to the simulated hits in order to reproduce the performance of the apparatus during data taking. The same analysis cuts as used for the data are applied to the tracks reconstructed from these hits.

The simulations (one for each resonance) are performed on a run-by-run basis, using a realistic description of the ALICE muon spectrometer performance. The misalignment of the muon spectrometer is tuned to reproduce the mass resolution of the $\jpsi$ measured from data. The resonances are generated in a $y$ range that is wider than the range used for the measurements ($2.5<y<4$) in order to account for edge effects. In each $y$ and $\pt$ interval, the $\aeffmean$ factor is calculated as the ratio of the number of reconstructed quarkonia over the number of quarkonia generated in this interval. 

The $\aeffmean$ factors, averaged over the entire data taking period, are $(13.22\pm0.02)\%$ for $\jpsi$, $(16.64\pm0.02)\%$ for $\psiprime$, $(20.93\pm0.05)\%$ for $\rm\Upsilon$(1S) and $(21.02\pm0.05)\%$ for $\rm\Upsilon$(2S), where the uncertainties are statistical. 
The $\aeffmean$ correction factors associated to the $\pt$ and $y$ differential yields are given in Section~\ref{sec:tables}.

\subsection{~\label{systematics}Systematic uncertainties}

The main sources of systematic uncertainties on the production cross section come from the estimation of the number of measured quarkonia, the acceptance times efficiency correction factor and the integrated luminosity. The uncertainty on the dimuon branching ratio is also taken into account.

The systematic uncertainty on the signal extraction is evaluated using the Root Mean Square (RMS) of the results obtained with different signal functions (CB2 or pseudo-Gaussian functions for charmonia, CB2 or Gaussian functions for bottomonia), different background functions (VWG or Pol4$\times$Exp for charmonia, the sum of two exponential or two power law functions for bottomonia) and different fitting ranges (beside the nominal fitting ranges quoted in Section~\ref{ssec:signal} the ranges $2.5<\mmumu<4.5$~GeV/$c^2$ and $8<\mmumu<12$~GeV/$c^2$ were also used for charmonia and bottomonia, respectively). The tail parameters of the signal functions are also varied within the limits determined by fits to the simulated quarkonium mass distributions in the $\pt$ or $y$ intervals used in the analysis. Finally, for the quarkonia analysis, different values for the ratio between the $\psiprime$ and the $\jpsi$ mass resolution have also been tested, estimated using a fit to the $\pt$- and $y$-integrated invariant mass distribution with these parameters left free. The resulting systematic uncertainties averaged over $\pt$ and $y$ are $2$\% for the $\jpsi$, $8$\% for the $\psiprime$, $8$\% for the $\rm\Upsilon$(1S) and $9$\% for the $\rm\Upsilon$(2S). 

The systematic uncertainty on the acceptance times efficiency correction factor has several contributions: the parametrization of the input $\pt$ and $y$ distributions of the simulated quarkonia, the track reconstruction efficiency, the trigger efficiency and the matching between tracks in the muon tracking and triggering chambers. The acceptance times efficiency correction factors are evaluated assuming that all quarkonium states are unpolarized. If the $\rm\Upsilon$(1S) production polarization is fully transverse or fully longitudinal, then the cross section changes by about 37\% and 20\%, respectively. This result is consistent with previous studies made for charmonia~\cite{Aamodt:2011gj,Aamodt:2011gjE}. There is to date no evidence for a significant quarkonium polarization at $\sqrt{s}=7$~TeV, neither for $\jpsi$~\cite{alice:2012pol}, $\psiprime$~\cite{Aaij:2013nlm,Chatrchyan:2013cla}, nor for $\rm\Upsilon$~\cite{Chatrchyan:2012woa}. Therefore, no systematic uncertainty due to the quarkonium polarization has been taken into account.

For $\jpsi$ and $\psiprime$, the parametrization of the input $\pt$ and $y$ distributions is based on fits to existing data measured at the same energy and in the same rapidity range~\cite{Aaij:2011jh,Aaij:2012ag}. The corresponding systematic uncertainty is obtained by varying these parametrizations within the statistical and systematic uncertainties of the data, and taking the RMS of the resulting $\aeffmean$ distribution.
Correlations between $\pt$ and $y$ observed by the LHCb collaboration~\cite{Aaij:2012ag} are also accounted for by evaluating the $\aeffmean$ factors for each $\pt$ ($y$) distribution measured in smaller $y$ ($\pt$) intervals and using the largest difference between the resulting values as an additional systematic uncertainty, quadratically summed to the one obtained using the procedure described above.
For the $\rm\Upsilon$, simulations are based on $\pt$ and $y$ parametrizations scaled from data measured by CDF~\cite{Abe:1995an} to $\sqrt{s}=7$~TeV. The corresponding systematic uncertainty is evaluated by changing the energy of the scaled CDF data to $\sqrt{s}=4$~TeV and $\sqrt{s}=10$~TeV and evaluating the corresponding $\aeffmean$. This corresponds to a variation of the input yields of at most 15\% as a function of rapidity and 40\% as a function of $\pt$. We note that extrapolating results obtained at a different collision energy is a conservative approach with respect to using CMS~\cite{Khachatryan:2010zg,Chatrchyan:2013yna} and LHCb~\cite{LHCb:2012aa} data at $\sqrt{s}=7$~TeV. The resulting uncertainties are 1.7\% for $\jpsi$ and $\psiprime$, and 2.4\% for $\rm\Upsilon$(1S) and $\rm\Upsilon$(2S). 

The single muon tracking efficiency can be evaluated both in data~\cite{Aamodt:2011gj} and in simulations. A difference of about $1.6$\% is observed which varies as a function of the muon pseudo-rapidity and $\pt$. The impact of this difference on $\aeffmean$ is quantified by replacing the single muon tracking efficiencies obtained from the simulated detector response with the values measured in the data. An additional uncertainty arising from the correlated inefficiency in the tracking chambers was evaluated and amounts to 2.5\% at the dimuon level. The resulting uncertainty on the corrected quarkonium yields amounts to 6.5\% for all resonances.

Concerning the trigger efficiency, a small difference is observed between data and simulations for the trigger response function. To account for this difference, a procedure similar to the one used for the systematic uncertainty on the track reconstruction efficiency is applied. The effect on $\aeffmean$ amounts up to $2$\% for all resonances. Additional uncertainties come from the method used to determine the RPC efficiency from data ($2$\%) and from the efficiency of the MB trigger condition for events where a quarkonium is produced ($2$\%). The latter uncertainty is evaluated by means of a sample of events collected with a stand-alone dimuon
trigger (without MB condition): the difference between the number of quarkonia reconstructed in such sample with and without the offline requirement of the MB condition is retained as uncertainty.

The difference observed in the simulations for different $\chi^2$ cuts on the matching between the tracks reconstructed in the tracking chambers and those reconstructed in the trigger chambers leads to a systematic uncertainty of 1\% on $\aeffmean$, independent from $\pt$ and $y$.

Finally, the uncertainty on the integrated luminosity amounts to 5\%. It includes contributions from the MB trigger cross section (3.5\%~\cite{Abelev:2012sea}), the MB trigger purity (3\%, evaluated by varying the cuts defining the beam-beam and beam-gas collisions), possible effects on the MB trigger cross section from V0 aging between the moment when the vdM scan was performed and the data taking period (1.5\%), the effects of V0 after-pulses and other instrumental effects on the MB trigger counts (1.5\%, evaluated from fluctuations in the ratio of the MB trigger rate to a reference trigger rate provided by the T0).

A summary of the different systematic sources is given in Table~\ref{tab-AccEff_syst} and the systematic uncertainties associated to the $\pt$ and $y$ differential cross sections are listed in Section~\ref{sec:tables}. Concerning the $\pt$ and $y$ dependence of these systematic uncertainties, the uncertainty associated to the luminosity is considered a global scale uncertainty, as is the uncertainty of the quarkonia branching ratio to dimuons. The one associated to the input MC parametrization is considered as largely point-to-point correlated. All other sources are considered as predominently uncorrelated.

\begin{table}[h] {
\begin{center} 
\begin{tabular}{cccccc}  
\hline
Source & $\jpsi$ & $\psiprime$ & $\rm\Upsilon$(1S) & $\rm\Upsilon$(2S) \\ 
\hline  Luminosity & 5\% & 5\% & 5\% & 5\% \\
Signal extraction & 2\% (2\%-15\%) & 8\% (7.5\%-11\%) & 8\% (8\%-13\%)& 9\% \\ 
Input MC parametrization & 1.7\% (0.1\%-1.8\%) & 1.7\% (0.4\%-2.4\%) & 2.4\% (0.6\%-4.5\%) & 2.4\% \\ 
Trigger efficiency & 3.5\% (3\%-5\%) & 3.5\% (3\%-5\%) & 3\% & 3\% \\ 
Tracking efficiency & 6.5\% (4.5\%-11.5\%) & 6.5\% (4.5\%-11.5\%) & 6.5\% (5.1\%-10.5\%)& 6.5\% \\ 
Tracking-trigger matching & 1\% & 1\% & 1\% & 1\% \\ 
\hline
\end{tabular}
\end{center}}
\caption{Relative systematic uncertainties on the quantities associated to quarkonium cross section measurement. Into brackets, values correspond to the minimum and the maximum as a function of $\pt$ and $y$.} 
\label{tab-AccEff_syst}
\end{table}

\section{\label{sec:results}Results}

\subsection{Integrated and differential production cross sections of $\jpsi$ and $\psiprime$}

The measured inclusive $\jpsi$ and $\psiprime$ production cross sections in $\pp$ collisions at $\sqrt{s}=7$~TeV in the rapidity range $2.5<y<4$ are:

$\sigma_{\rm \jpsi}=~6.69\pm0.04({\rm stat})\pm0.63({\rm syst})$~$\mu$b, for $0<\pt<20$~GeV$/c$,

$\sigma_\psiprime=1.13\pm0.07({\rm stat})\pm0.19({\rm syst})$~$\mu$b, for $0<\pt<12$~GeV$/c$.

The measured $\jpsi$ production cross section is in good agreement with the previously published ALICE result~\cite{Aamodt:2011gj,Aamodt:2011gjE}. 

\begin{figure}[h!]
\includegraphics[width=0.5\linewidth,keepaspectratio]{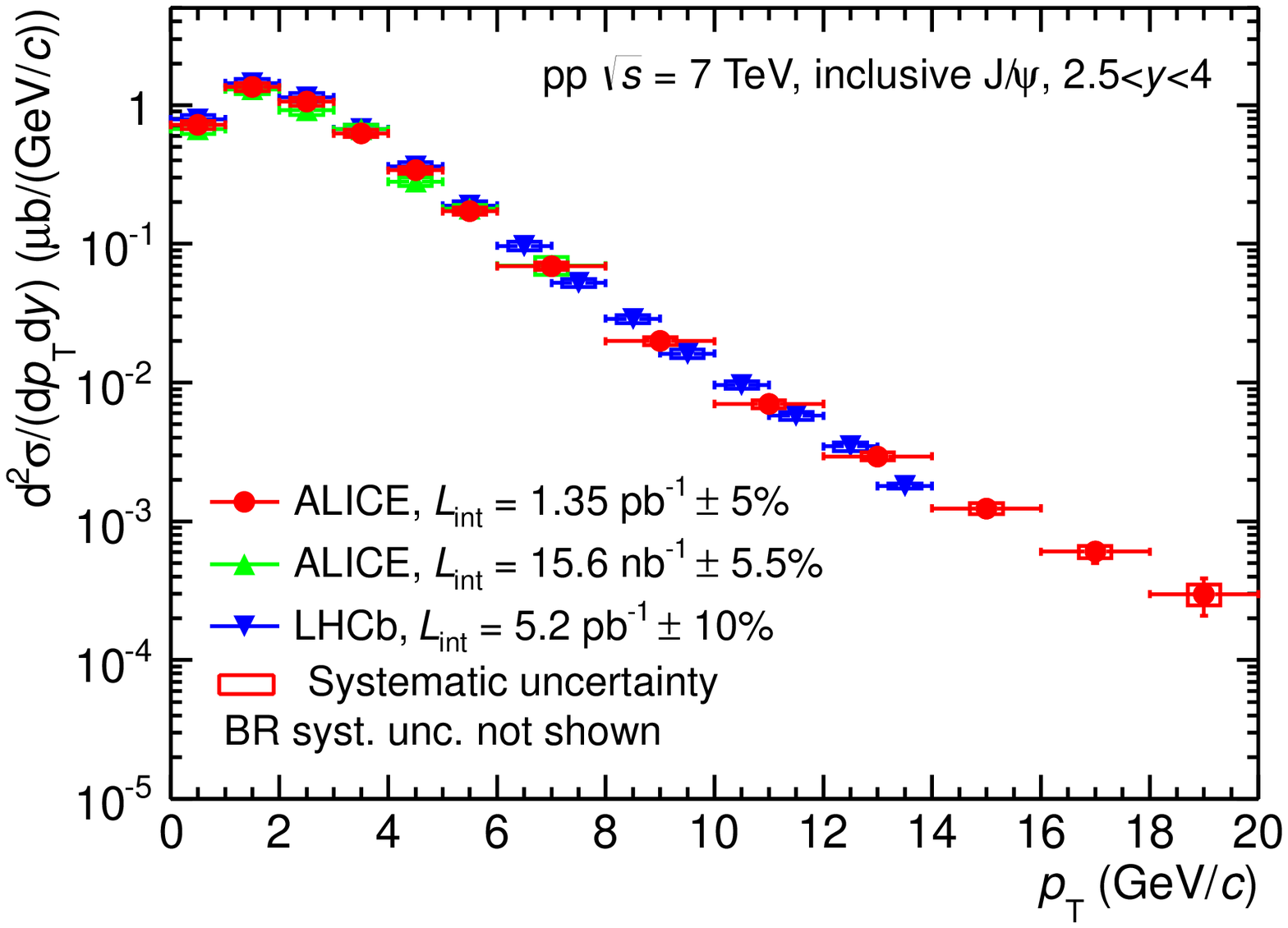}
\includegraphics[width=0.5\linewidth,keepaspectratio]{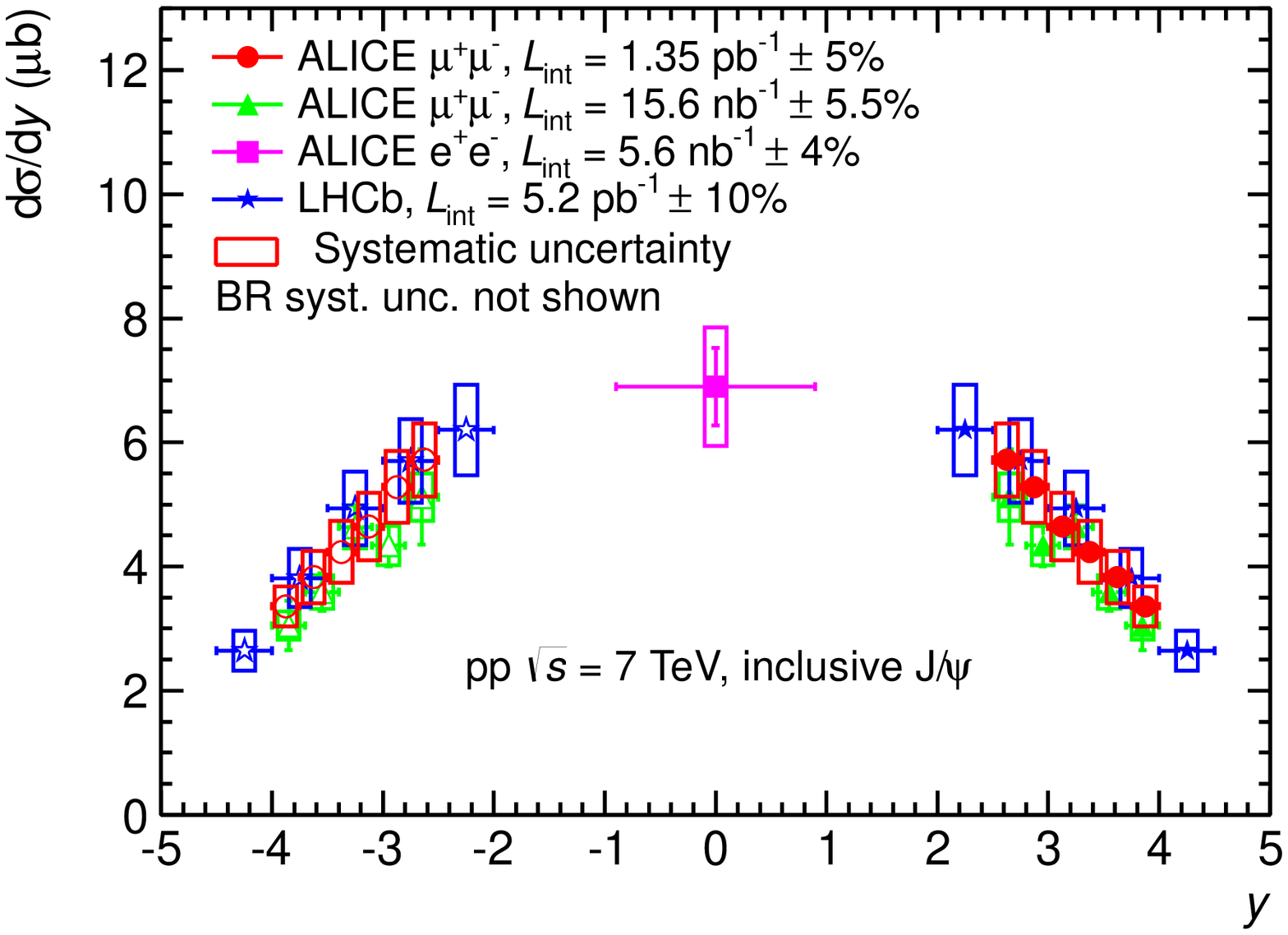}\\
\includegraphics[width=0.5\linewidth,keepaspectratio]{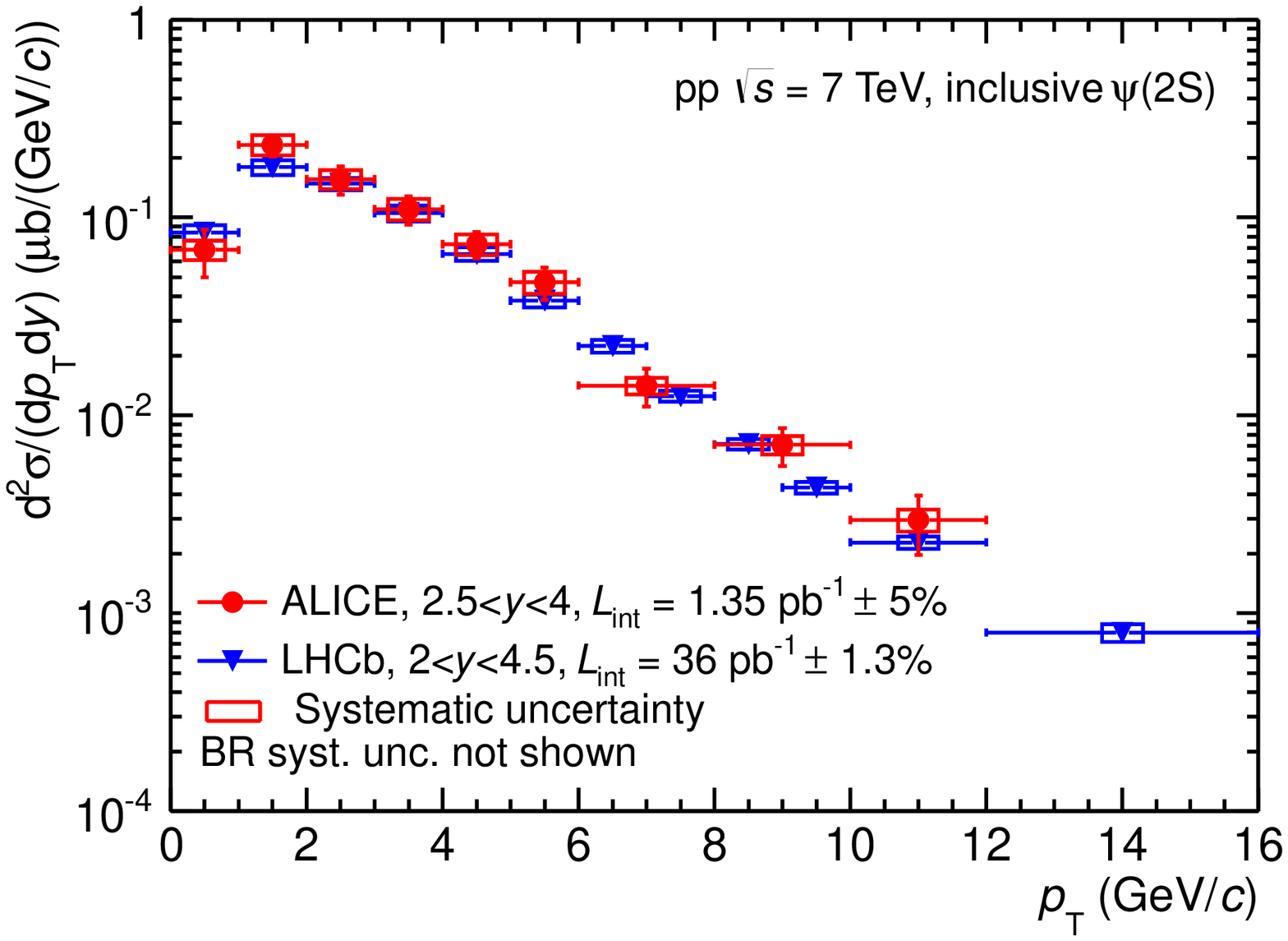}
\includegraphics[width=0.5\linewidth,keepaspectratio]{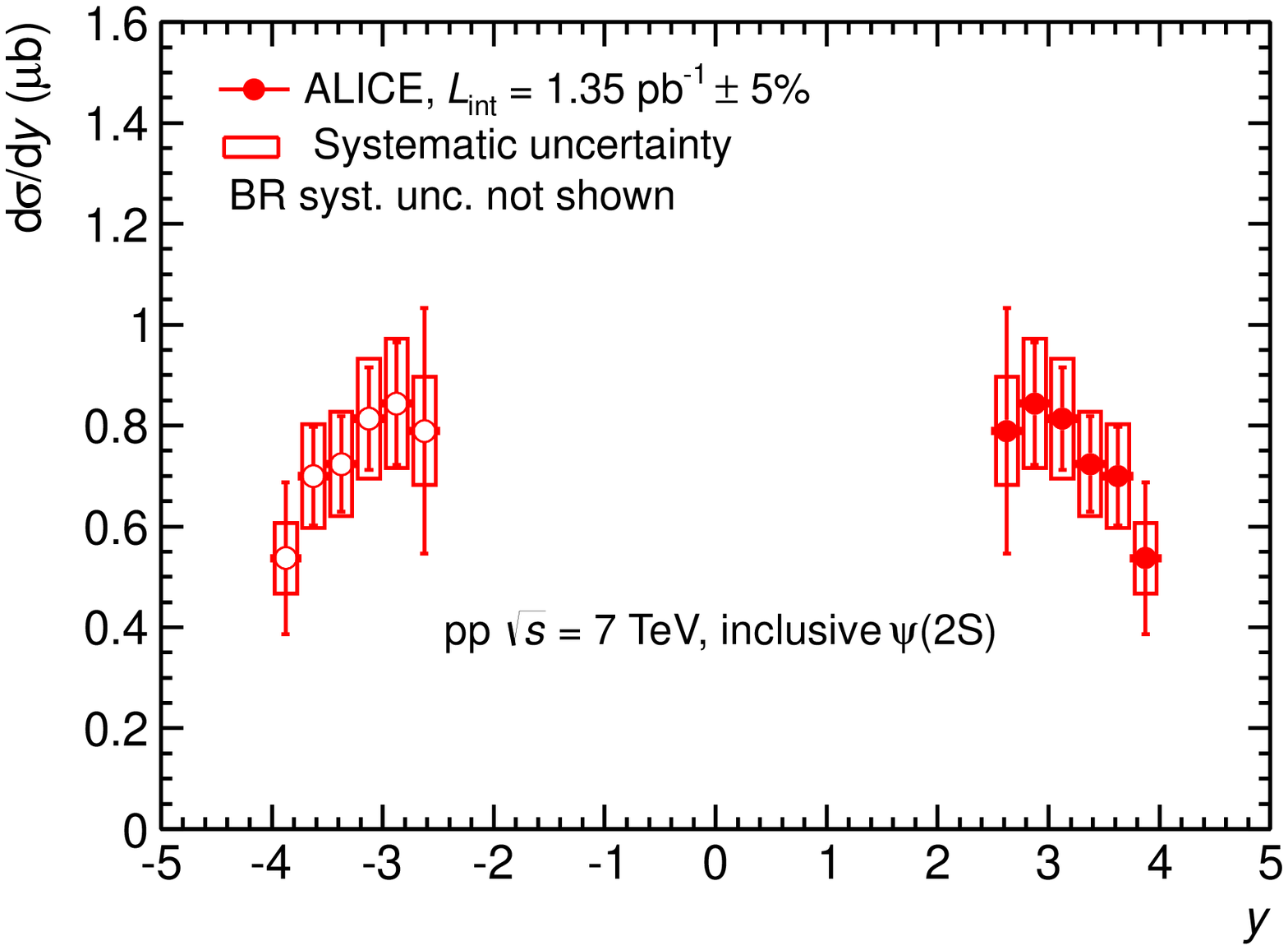}
\caption{Differential production cross sections of $\jpsi$ (top) and $\psiprime$ (bottom) as a function of $\pt$ (left) and $y$ (right). The results are compared to previous ALICE results~\cite{Aamodt:2011gj,Aamodt:2011gjE} and LHCb measurements~\cite{Aaij:2011jh,Aaij:2012ag}. The open symbols are the reflection of the positive-y measurements with respect to $y=0$. The vertical error bars and the boxes represent the statistical and systematic uncertainties, respectively.}
\label{cross_jpsi}
\end{figure}

Figure~\ref{cross_jpsi} shows the differential production cross sections of $\jpsi$ (top) and $\psiprime$ (bottom) as a function of $\pt$ (left) and rapidity (right). In all figures, the error bars represent the statistical uncertainties whereas the boxes correspond to the systematic uncertainties. The systematic uncertainty on the luminosity is quoted in the legend. This analysis extends the $\pt$ range of the $\jpsi$ measurement with respect to the previous ALICE measurement~\cite{Aamodt:2011gj,Aamodt:2011gjE} from 8 GeV/$c$ to 20~GeV/$c$. 

The $\pt$ differential cross sections are compared with the values reported by the LHCb collaboration~\cite{Aaij:2011jh,Aaij:2012ag}. The LHCb data points in Figure~\ref{cross_jpsi} correspond to the sum of prompt and $b$-meson decays quarkonium productions. For the $\jpsi$ cross sections (Fig.~\ref{cross_jpsi}, top left), a good agreement is observed between the two experiments. 
The comparison to the LHCb results for the $\pt$ dependence of $\psiprime$ cross section 
(Fig.~\ref{cross_jpsi}, bottom left) is not straightforward due to the different  rapidity ranges. The ALICE measurement tends to be slightly higher than the one reported by LHCb, except at very low $\pt$. Still, the results are in agreement within systematic uncertainties.

The differential cross sections of $\jpsi$ as a function of rapidity (Fig.~\ref{cross_jpsi}, top right) are compared to the previous measurements reported by ALICE \cite{Aamodt:2011gj,Aamodt:2011gjE} and LHCb \cite{Aaij:2011jh}. The results are in good agreement. Furthermore, the ALICE $\jpsi$ measurement at mid-rapidity in the di-electron channel complements the forward rapidity measurement and allows to present the $\jpsi$ differential cross section over a broad rapidity range for $\pt$ down to zero. The rapidity dependence of the inclusive $\psiprime$ production cross section at forward rapidity (Fig.~\ref{cross_jpsi}, bottom right) is measured for the first time at $\sqrt{s}=7$~TeV. 

The inclusive $\psiprime$-to-$\jpsi$ cross section ratio at $\sqrt{s}=7$~TeV, integrated over $\pt$ and $y$, is $\sigma_\psiprime/\sigma_\jpsi=0.170\pm0.011({\rm stat.})\pm0.013({\rm syst})$. To obtain this ratio, the same fit function (CB2 or pseudo-Gaussian function) is used for both resonances, for all the cases described in Section~\ref{systematics}. The mean of the resulting distribution is used as the central value and its RMS is used as the systematic uncertainty on signal extraction. The other sources of systematic uncertainty cancel out in the ratio, except for the uncertainty on the $\aeffmean$ factors. As a consequence of the adopted procedure, some differences between this value and the ratio of the integrated cross sections are expected.

\begin{figure}[h]
\includegraphics[width=0.5\textwidth]{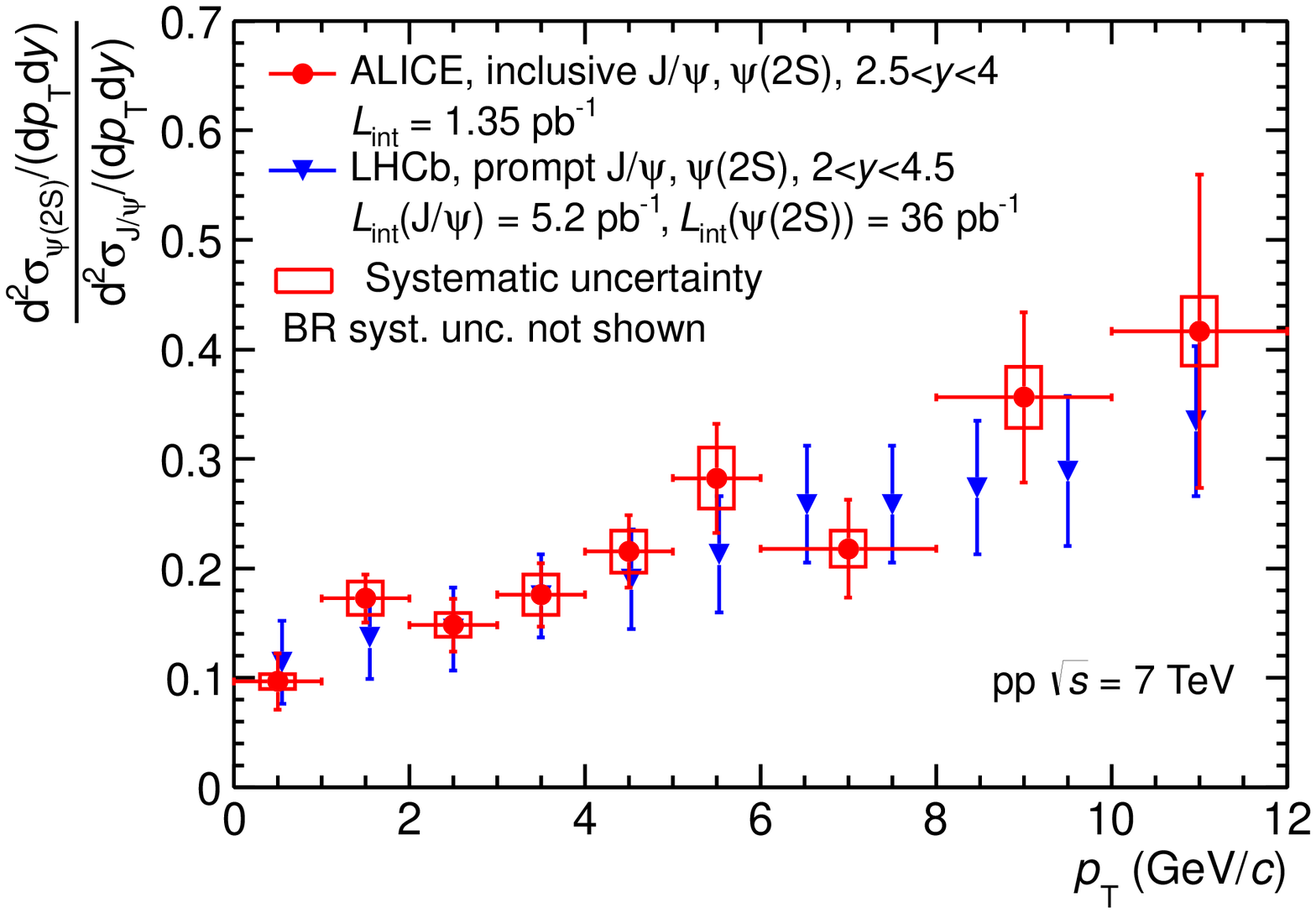}
\includegraphics[width=0.5\textwidth]{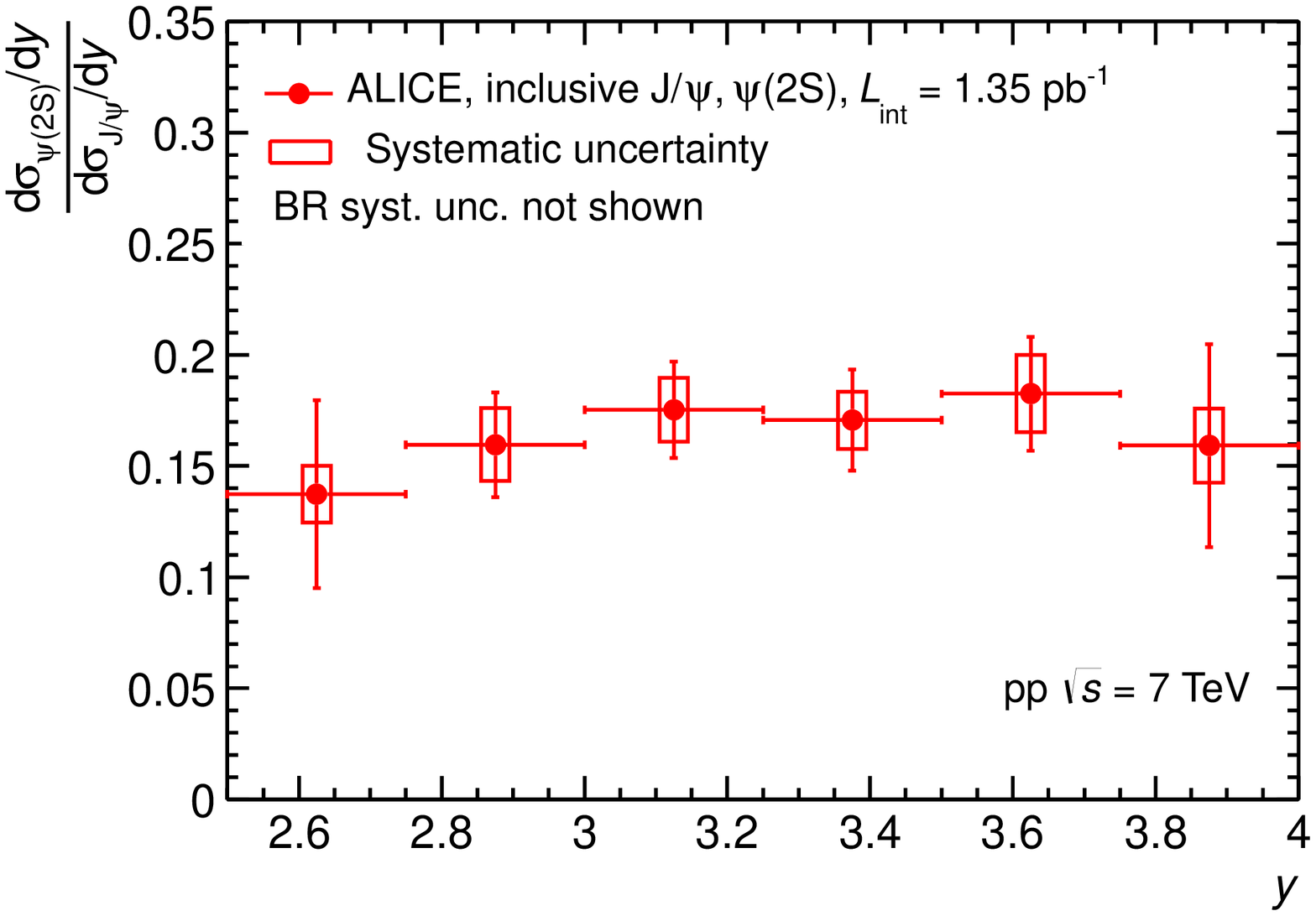}
\caption{$\psiprime$/$\jpsi$ ratio as a function of $\pt$ (left) compared to LHCb measurement~\cite{Aaij:2012ag} and as a function of rapidity (right).}
\label{ratio}
\end{figure}

Figure~\ref{ratio} presents the $\psiprime$-to-$\jpsi$ cross section ratio as a function of $\pt$ (left) and $y$ (right). This ratio increases with $\pt$, whereas it shows little or no dependence on rapidity. The comparison with the LHCb measurement (left) shows a reasonable agreement, even though this analysis presents the ratio between inclusive cross sections whereas the LHCb collaboration reports the ratio between prompt particle cross sections, thus removing the contribution from $b$-meson decays.
Assuming that the $\psiprime$-to-$\jpsi$ cross section ratio is independent of $y$ over the entire rapidity range, as confirmed by ALICE measurements, and multiplying it by the branching ratio of $\psiprime$ decaying into $\jpsi$ plus anything ${\rm BR}_{\psiprime\rightarrow\jpsi}=60.3\pm0.7$\%~\cite{PhysRevD.86.010001}, one gets the fraction of inclusive $\jpsi$ coming from $\psiprime$ decay $f^\psiprime=0.103\pm0.007({\rm stat})\pm0.008({\rm syst})$.

\subsection{Integrated and differential production cross sections of $\rm\Upsilon$(1S) and $\rm\Upsilon$(2S)}

The measured inclusive $\rm\Upsilon$(1S) and $\rm\Upsilon$(2S) production cross sections, integrated over $2.5 < y < 4$ and $0<\pt<12$~GeV/$c$, are: 

$\sigma_{\rm\Upsilon({\rm 1S})}=54.2\pm5.0 ({\rm stat})\pm6.7({\rm syst})$~nb 

$\sigma_{\rm\Upsilon({\rm 2S})}=18.4\pm3.7({\rm stat})\pm2.9({\rm syst})$~nb.

The total number of $\rm\Upsilon$(1S) extracted from the data allows to measure its differential production cross section in five $\pt$ intervals and three rapidity intervals. For the $\rm\Upsilon$(2S), on the contrary, no differential analysis could be performed due to the limited number of events.

\begin{figure}[!hbt]
\includegraphics[width=0.5\textwidth]{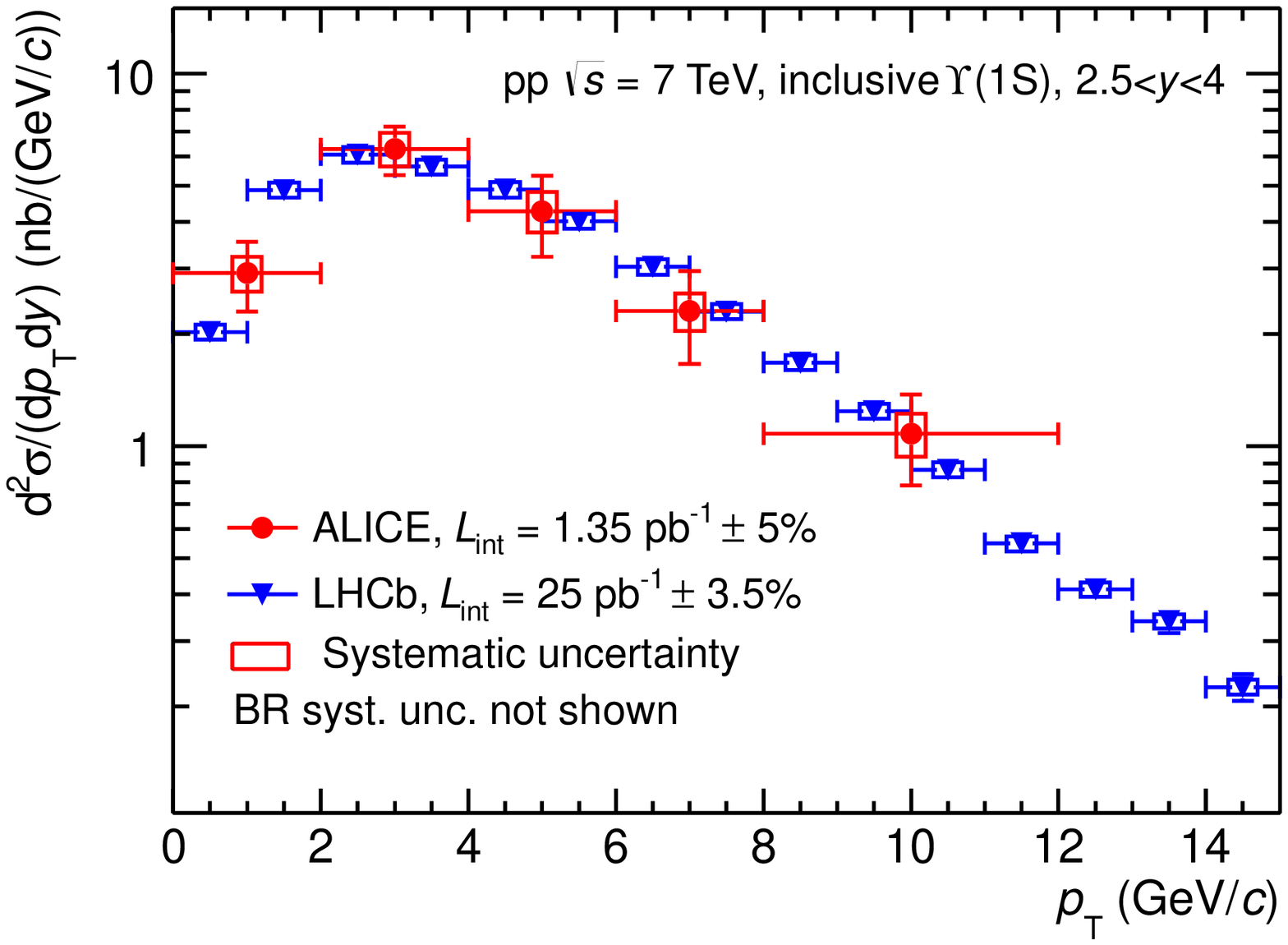}
\includegraphics[width=0.5\textwidth]{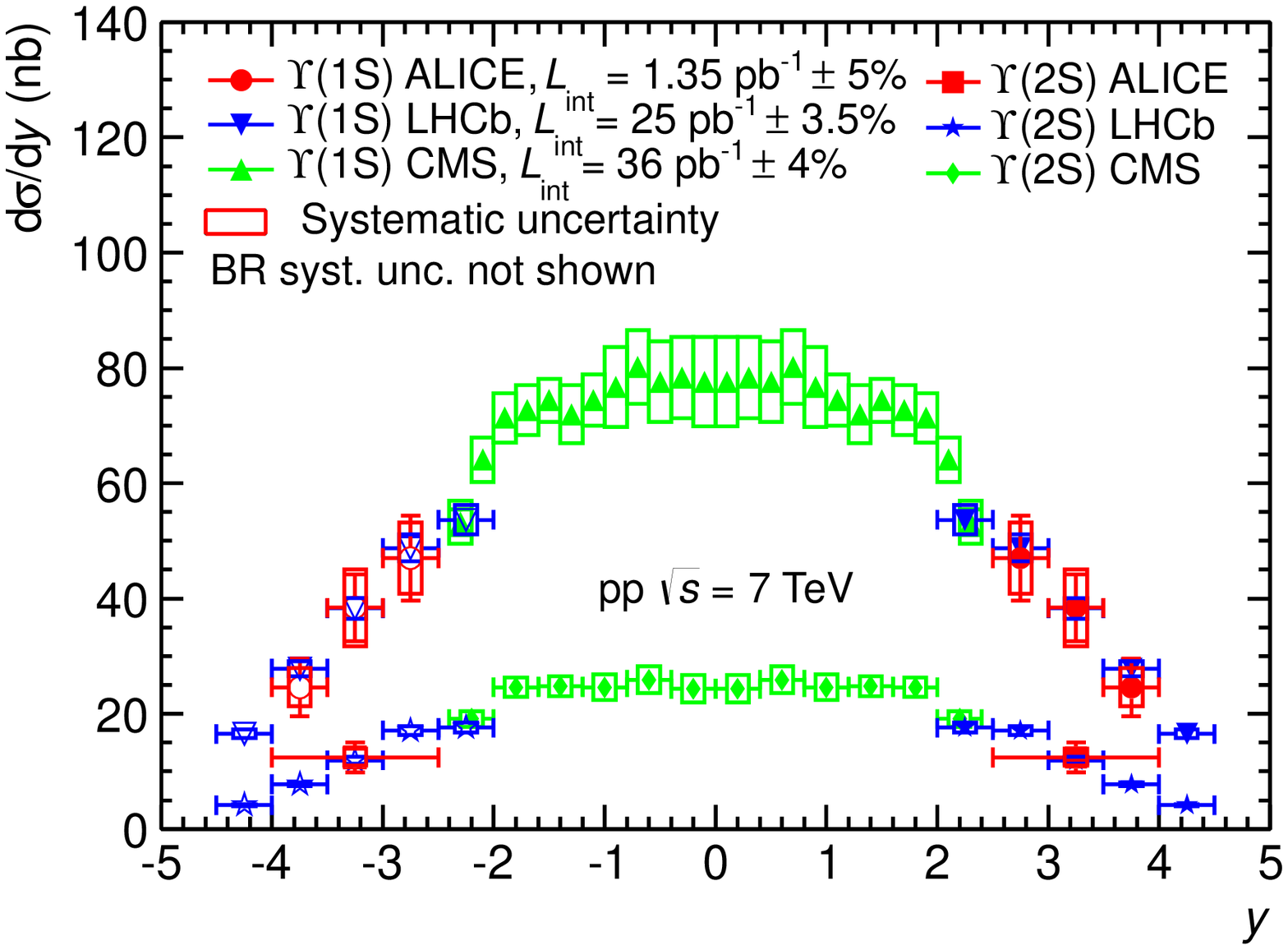}
\caption{Differential cross section of $\rm\Upsilon$(1S) as a function of $\pt$ (left) and differential cross sections of $\rm\Upsilon$(1S) and $\rm\Upsilon$(2S) as function of rapidity (right), measured by ALICE, LHCb~\cite{LHCb:2012aa} and CMS~\cite{Khachatryan:2010zg,Chatrchyan:2013yna}. The open symbols are reflected with respect to $y=0$.}
\label{fig_upsi3}
\end{figure}

Figure~\ref{fig_upsi3} presents the $\rm\Upsilon$(1S) differential production cross section as a function of $\pt$ (left) and the differential cross sections of $\rm\Upsilon$(1S) and $\rm\Upsilon$(2S) as a function of rapidity (right). The $\rm\Upsilon$(1S) $\pt$ differential cross sections are compared to the values reported by the LHCb collaboration~\cite{LHCb:2012aa} in the same rapidity range ($2.5 < y < 4$). The results are in good agreement. The $\rm\Upsilon$(1S) and $\rm\Upsilon$(2S) differential cross sections as a function of rapidity (Fig.~\ref{fig_upsi3} right) are presented together with the LHCb~\cite{LHCb:2012aa} and CMS ~\cite{Khachatryan:2010zg,Chatrchyan:2013yna} measurements for $\pt$ down to zero. The measurements from ALICE and LHCb are in good agreement for both $\rm\Upsilon$ states.

The $\rm\Upsilon$(2S)-to-$\rm\Upsilon$(1S) cross section ratio at $\sqrt{s}=7$~TeV integrated over $\pt$ and $y$ is: $\sigma_{\rm\Upsilon({\rm 2S})}/\sigma_{\rm\Upsilon({\rm 1S})}=0.34\pm0.10({\rm stat})\pm0.02({\rm syst})$. This ratio is in agreement with the one measured by the LHCb experiment~\cite{LHCb:2012aa}. Using a branching ratio for $\rm\Upsilon$(2S) decaying into $\rm\Upsilon$(1S) plus anything  ${\rm BR}_{{\rm\Upsilon(2S)}\rightarrow{\rm\Upsilon(1S)}}=26.5\pm0.5$\%~\cite{PhysRevD.86.010001}, one gets the fraction of inclusive $\rm\Upsilon$(1S) coming from $\rm\Upsilon$(2S) decay $f^{\rm\Upsilon({\rm 2S})}=0.090\pm0.027({\rm stat})\pm0.005({\rm syst})$. 

\section{\label{sec:theory}Model comparison}

\subsection{\label{subsec:theory_pt}Differential production cross sections as a function of $\pt$}

\begin{figure}[h!]
\centerline{\includegraphics[width=0.5\linewidth,keepaspectratio]{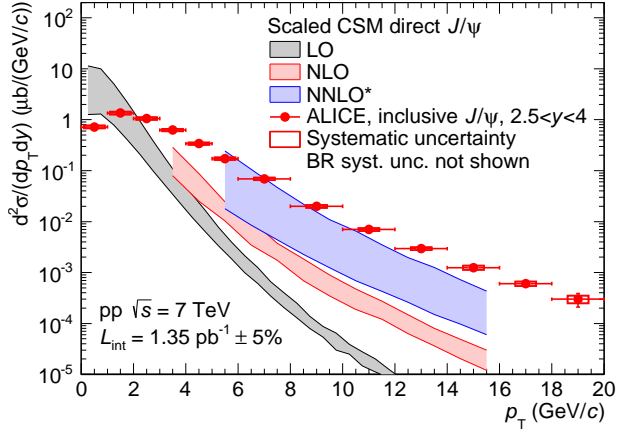}}
\caption{(color online). Inclusive $\jpsi$ differential production cross section as a function of $\pt$, compared to several scaled CSM calculations for direct $\jpsi$~\cite{Lansberg:2011hi}. Details on the calculations are given in the text.}
\label{jpsi_theory_pt_csm}
\end{figure}

The measured inclusive $\jpsi$ differential production cross section as a function of $\pt$ is compared to three theoretical calculations performed in the CSM (Fig.~\ref{jpsi_theory_pt_csm}): two complete calculations at LO and NLO respectively and a third calculation, called NNLO*, that includes the leading-$\pt$ contributions appearing at NNLO~\cite{Lansberg:2011hi}. 
In agreement with the authors, the calculations are scaled by a factor $1/0.6$ to account for the fact that they correspond to direct $\jpsi$ production, whereas they are compared to inclusive measurements. This scaling factor is obtained by assuming that about $20$\% of the inclusive $\jpsi$ come from $\chic$ decay~\cite{LHCb:2012af}, $10$\% from $\psiprime$ (factor $f^\psiprime$, Section~\ref{sec:results}) and $9$\% from $b$-mesons~\cite{Aaij:2011jh}. The LO calculation underestimates the data for $\pt>2$~GeV/$c$ and the $\pt$ dependence is much steeper than the measured one. At NLO, the $\pt$ dependence is closer to that of the data, but the calculation still underestimates the measured cross section. The addition of some NNLO contributions further improves the agreement between data and theory concerning the $\pt$ dependence and further reduces the difference between the two, at the price of larger theoretical uncertainties. 

Using a constant scaling factor for the direct-to-inclusive $\jpsi$ production cross section ratio requires that the $\pt$ distributions of direct and decay $\jpsi$ have the same shape. This assumption is a rather crude approximation and for instance the LHCb collaboration has measured a significant increase of the fraction of $\jpsi$ from $b$-meson decay with $\pt$ up to $30$\% for $\pt>14$~GeV/$c$~\cite{Aaij:2011jh}. Properly accounting for these variations would improve the agreement between data and theory at large $\pt$.

\begin{figure}[h!]
\begin{tabular}{cc}
\includegraphics[width=0.5\linewidth,keepaspectratio]{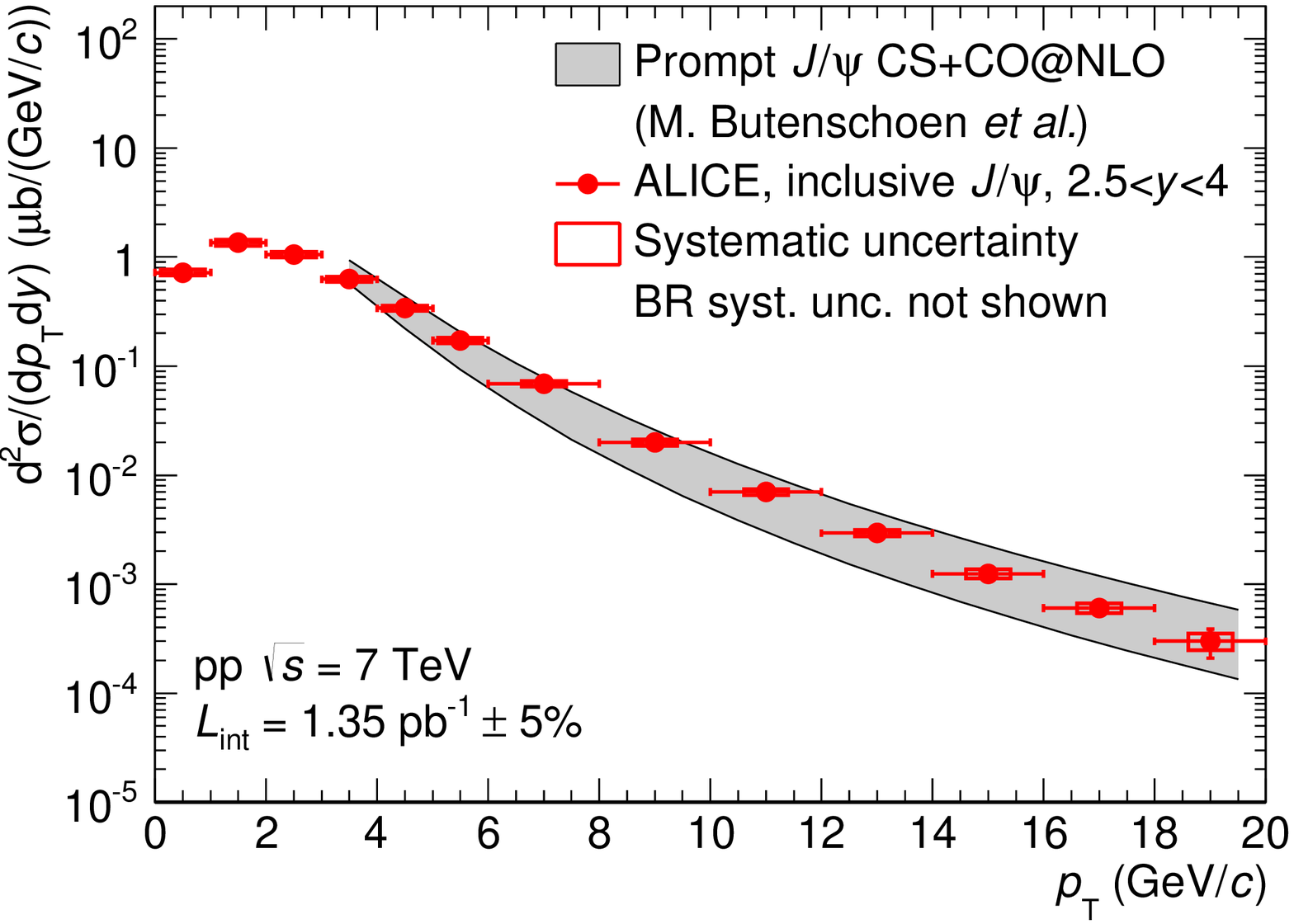}&
\includegraphics[width=0.5\linewidth,keepaspectratio]{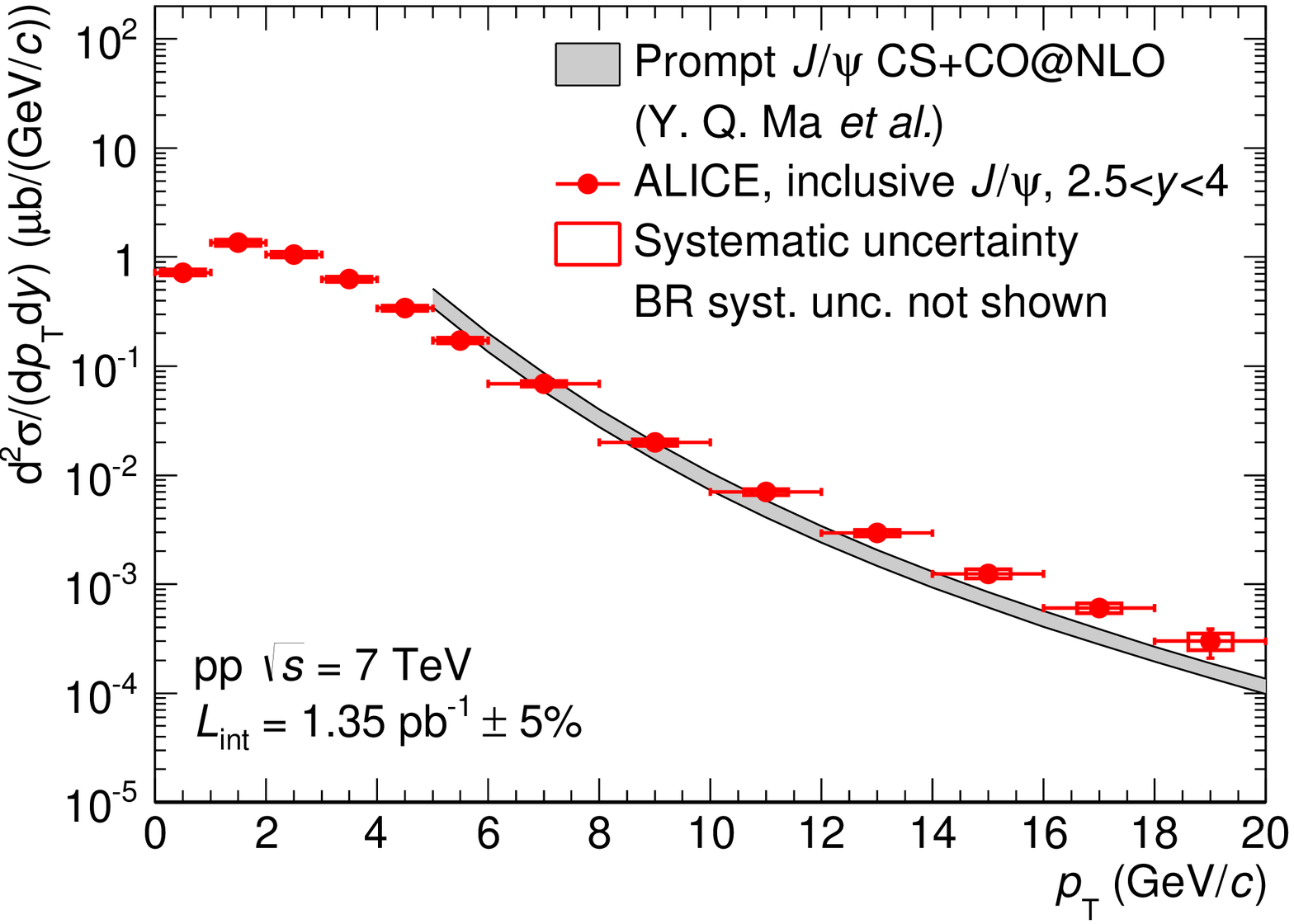}\\
\includegraphics[width=0.5\linewidth,keepaspectratio]{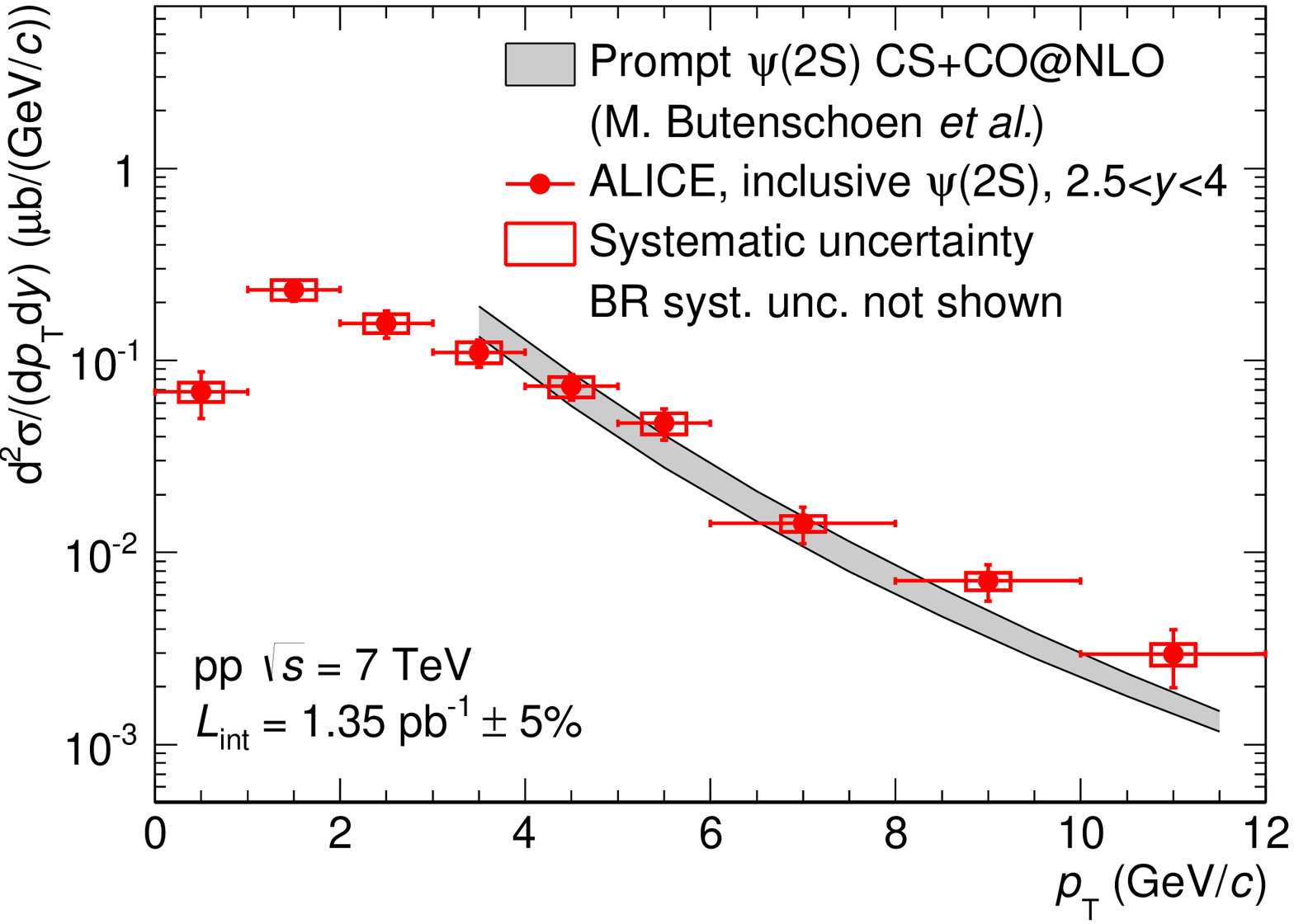}&
\includegraphics[width=0.5\linewidth,keepaspectratio]{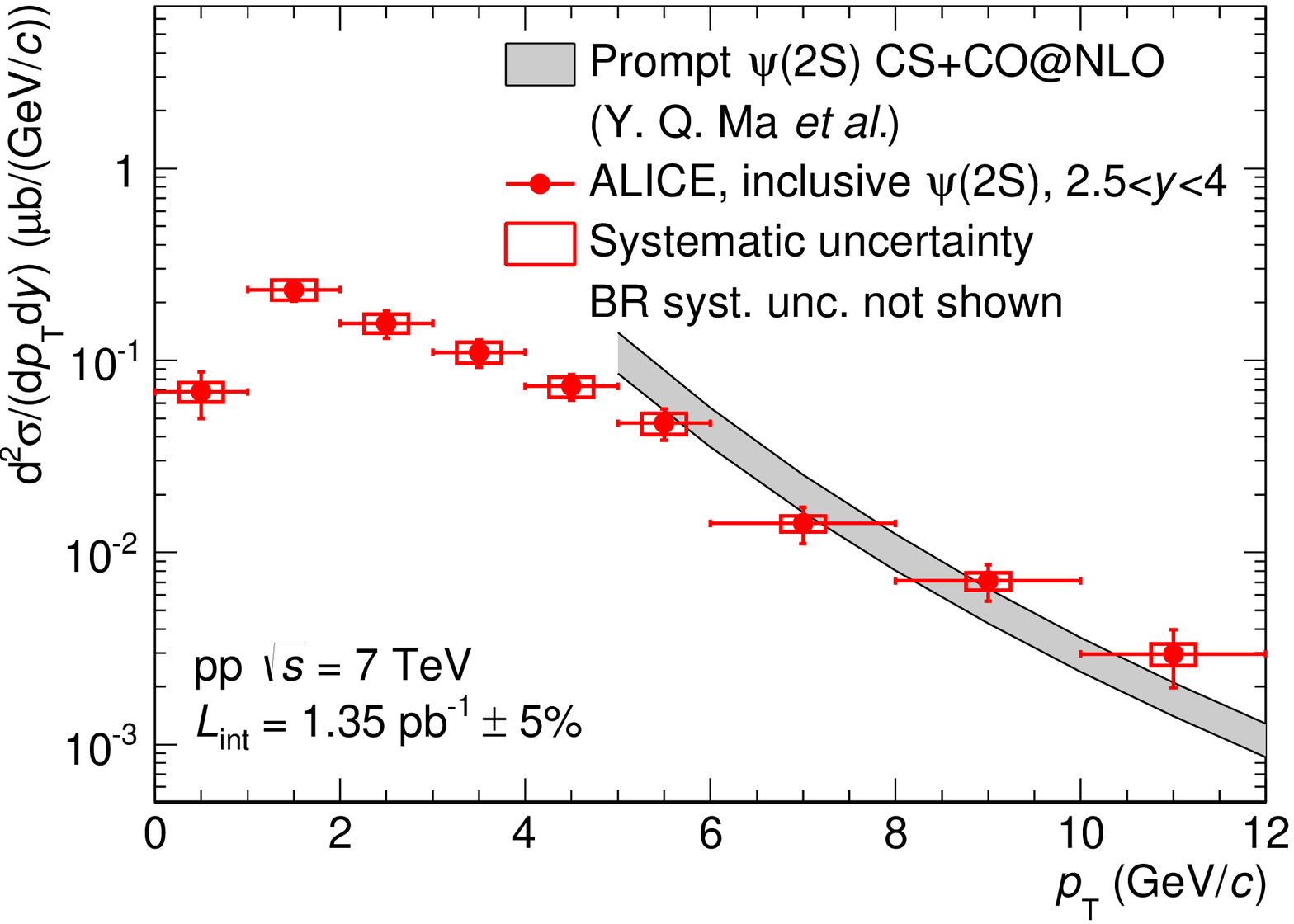}\\
\includegraphics[width=0.5\linewidth,keepaspectratio]{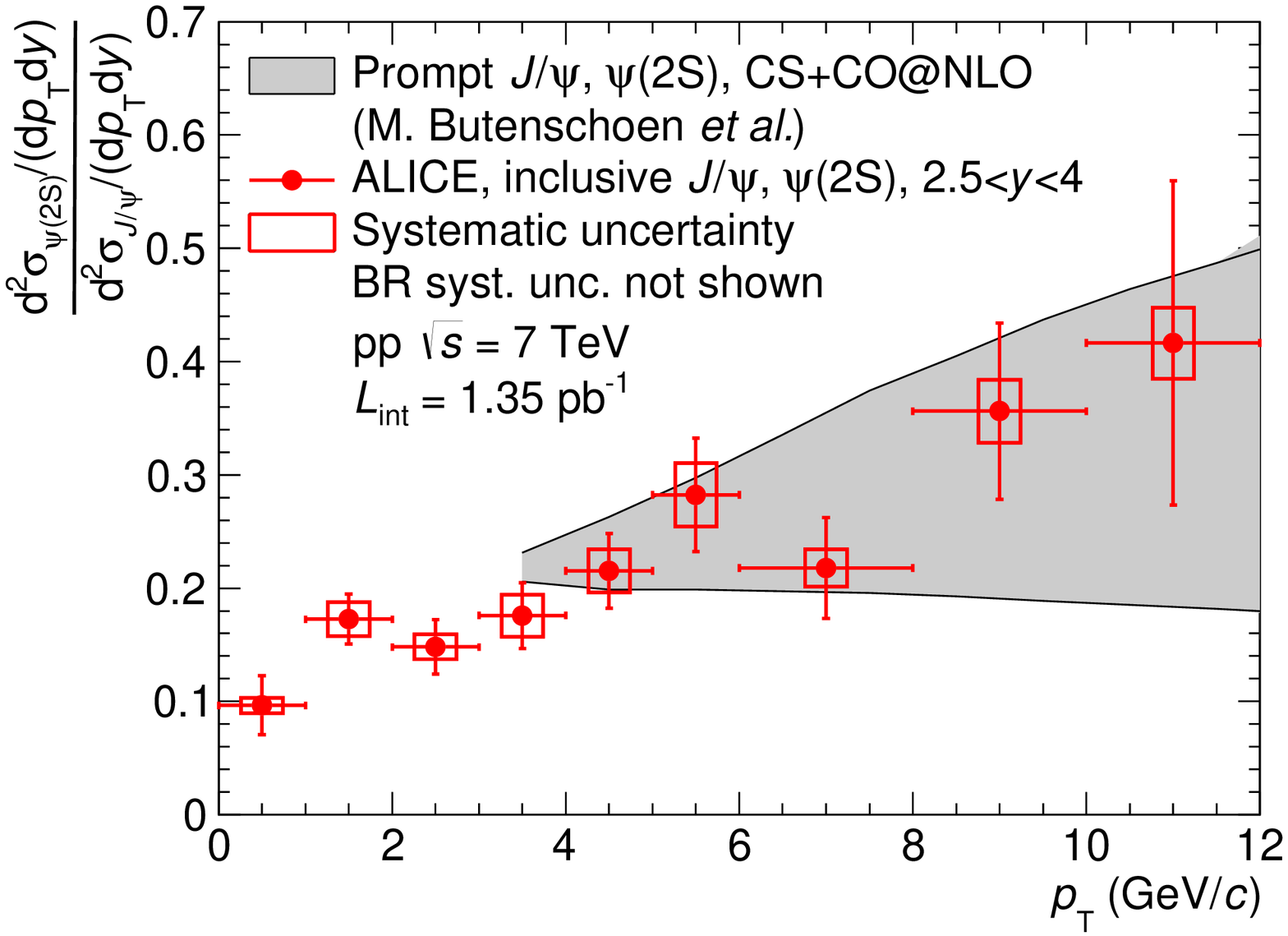}&
\includegraphics[width=0.5\linewidth,keepaspectratio]{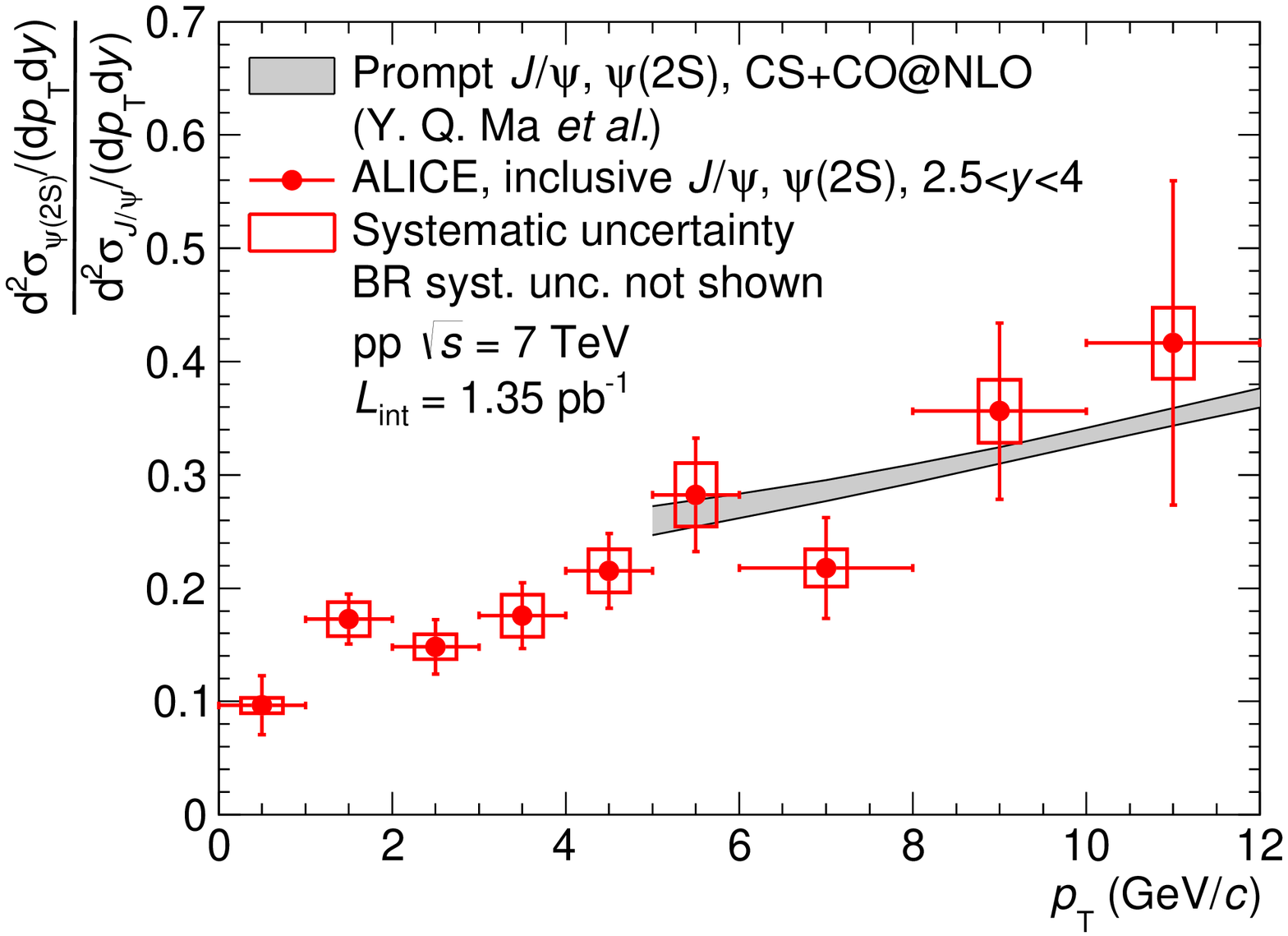}
\end{tabular}
\caption{Inclusive $\jpsi$ differential production cross section (top), inclusive $\psiprime$ differential production cross section (middle) and inclusive $\psiprime$-to-$\jpsi$ ratio (bottom) as a function of $\pt$ compared to two NRQCD calculations from~\cite{Butenschoen:2011yh} (left) and \cite{Ma:2010jj} (right).}
\label{psi_theory_pt_com}
\end{figure}

Figure~\ref{psi_theory_pt_com} presents the comparison of the inclusive $\jpsi$ differential production cross section (top), the inclusive $\psiprime$ differential production cross section (middle) and the ratio between the two (bottom) as a function of $\pt$ to two NRQCD calculations for prompt $\jpsi$ and $\psiprime$ production at NLO from~\cite{Butenschoen:2011yh} (left) and \cite{Ma:2010jj} (right). As discussed with the authors, a number of theoretical uncertainties cancels out when forming the $\psiprime$-to-$\jpsi$ ratio and the theory bands shown in the bottom panels are obtained by taking the ratio of the $\psiprime$ and $\jpsi$ upper and lower bounds from top and middle panels separately, rather than forming all four combinations.

The NRQCD calculations include both the same leading order Color-Singlet (CS) contributions as the one shown in Fig.~\ref{jpsi_theory_pt_csm} and Color-Octet (CO) contributions that are adjusted to experimental data by means of so-called Long-Range Matrix Elements (LRME). The two calculations differ in the LRME parametrization: the first (left panels of Fig.~\ref{psi_theory_pt_com}) uses three matrix elements whereas the second (right panels of Fig.~\ref{psi_theory_pt_com}) uses only two linear combinations of these three elements. Other differences include: the data sets used to fit these matrix elements, the minimum $\pt$ above which the calculation is applicable and the way by which contributions from $\chic$ decays into prompt $\jpsi$ and $\psiprime$ productions are accounted for. The first calculation has significantly larger uncertainties than the second for both the $\jpsi$ cross section and the $\psiprime$-to-$\jpsi$ ratio. This is a consequence of the differences detailed above and in particular the fact that the fits start at a lower $\pt$ and include a larger number of data sets.

Both calculations show reasonable agreement with data for all three observables. As it is the case for the CSM calculations, properly accounting for the contribution from $b$-meson decays to both $\jpsi$ and $\psiprime$ inclusive productions in either the data or the theory would further improve the agreement at high $\pt$. 

In the CSM, the direct $\psiprime$ to direct $\jpsi$ ratio is a constant, independent of $\pt$ and rapidity.
It corresponds to the square of the ratio between the $\psiprime$ and $\jpsi$ wave functions at the origin and amounts to about $0.6$~\cite{Lansberg:2011hi}\footnote{There is no uncertainty on this quantity because none is quoted for the $\psiprime$ wave function taken from~\cite{Eichten:1995ch}.}.
This value, scaled by the direct-to-inclusive $\jpsi$ and $\psiprime$ ratios ($0.6$ for $\jpsi$, as discussed above, and $0.85$ for $\psiprime$~\cite{Aaij:2012ag}), becomes 0.42.
It is larger than the $\pt$-integrated measurement quoted in Section~\ref{sec:results} and matches the values measured for $\pt>9$~GeV/$c$.

\begin{figure}[h!]
\begin{center}
\includegraphics[width=0.5\linewidth,keepaspectratio]{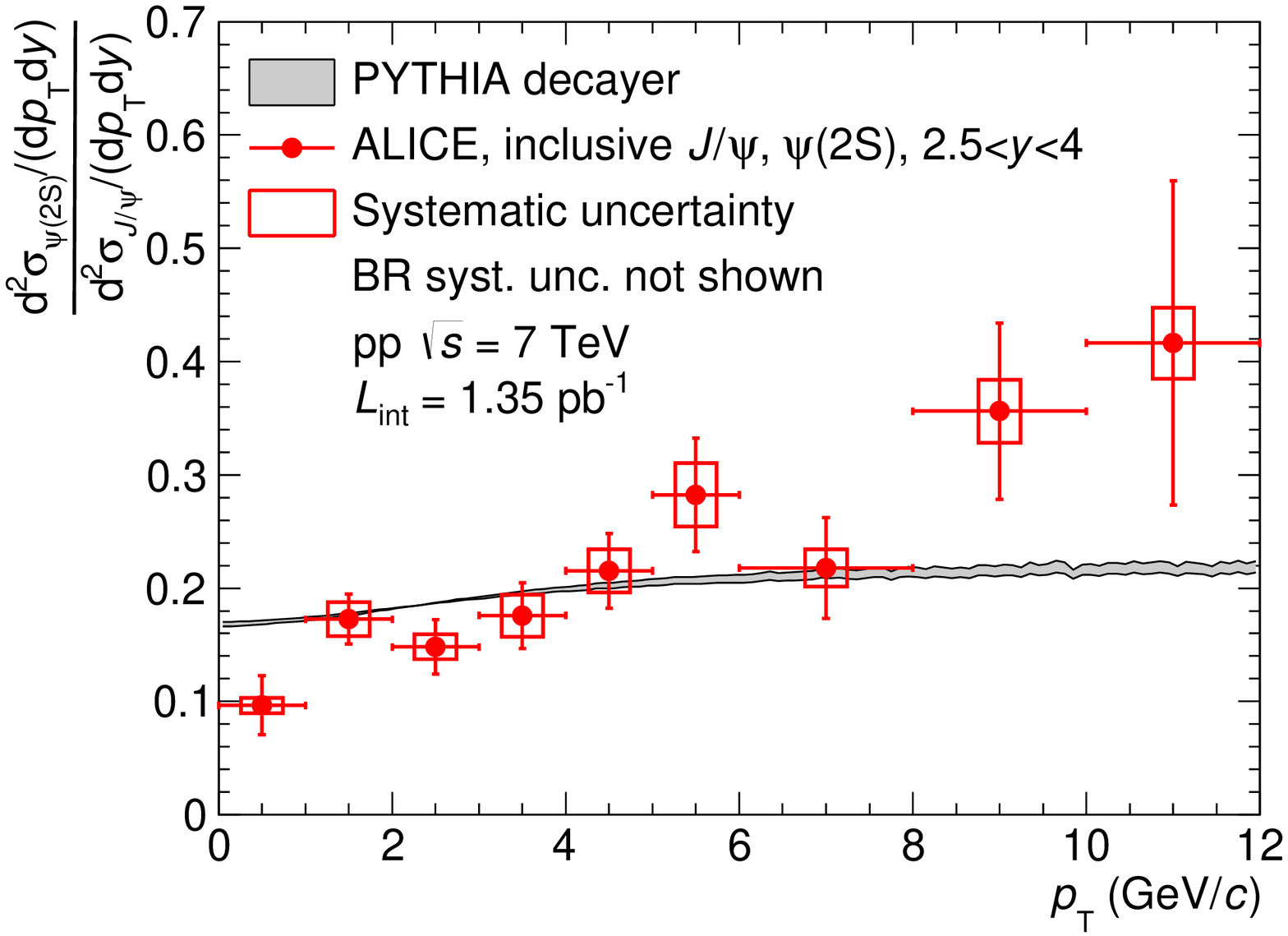}
\end{center}
\caption{Inclusive $\psiprime$-to-$\jpsi$ cross section ratio as a function of $\pt$ compared to a simulation in which all direct quarkonia are considered to have the same $\pt$ distribution and only kinematic effects due to the decay of higher mass resonances are taken into account, using PYTHIA~\cite{pythia}.} 
\label{pythia_theory_pt}
\end{figure}

Concerning the increase of the inclusive $\psiprime$-to-$\jpsi$ cross section ratio as a function of $\pt$ observed in the data, a fraction originates from the contribution of $\psiprime$ and $\chic$ decays. Assuming that the direct production of all charmonium states follows the same $\pt$ distribution, as it is the case in the CEM, the transverse momentum of $\jpsi$ coming from the decay of the higher mass resonances must be smaller than the one of the parent particle. This results in an increase of the corresponding contribution to the inclusive cross section ratio as a function of $\pt$. The $\pt$ dependence resulting from this effect on the inclusive $\psiprime$-to-$\jpsi$ cross section ratio has been investigated using PYTHIA~\cite{pythia} for decaying the 
parent particle into a $\jpsi$. The result is normalized to our measured integrated $\psiprime$-to-$\jpsi$ cross section ratio and compared to the data in Fig.~\ref{pythia_theory_pt}.
As expected, an increase of the ratio is observed with increasing $\pt$ but it is not sufficient to explain the trend observed in the data. 
This indicates that the increase observed in the data cannot be entirely explained with simple decay kinematics arguments and that other effects must be taken into account. 
A non-constant ratio can already be expected in the simplest case of CSM, where different diagram contributions to S- and P- wave charmonia production are expected, resulting in different feed-down contributions to $\jpsi$ and $\psiprime$.
On top of this Color-Octet contributions can also be added, as done in the NRQCD framework. The proper accounting of such contributions is sufficient to reproduce the trend observed in the data, as shown in Fig.~\ref{psi_theory_pt_com}, bottom panels.

\begin{figure}[h!]
\includegraphics[width=0.5\linewidth,keepaspectratio]{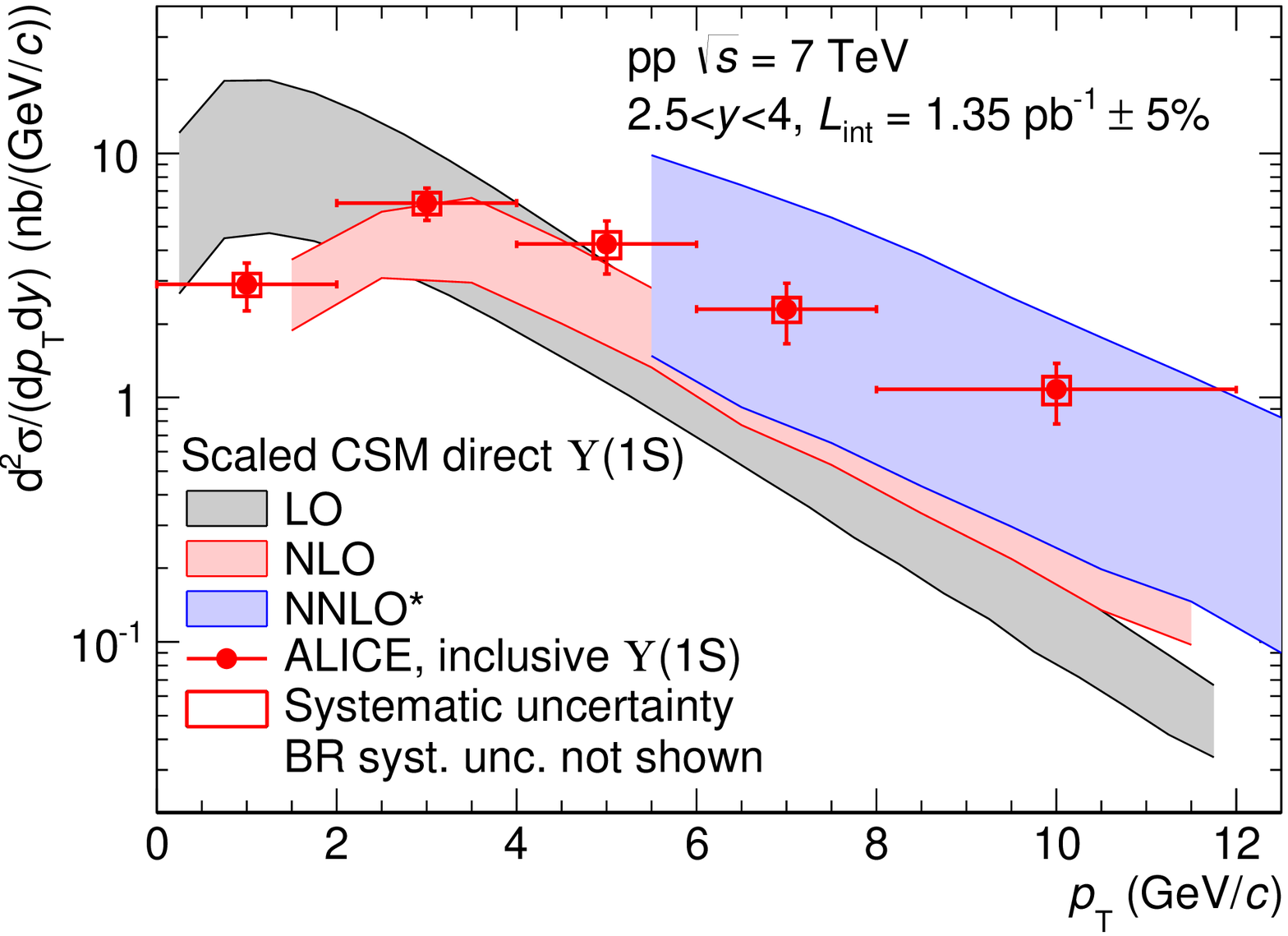}
\includegraphics[width=0.5\linewidth,keepaspectratio]{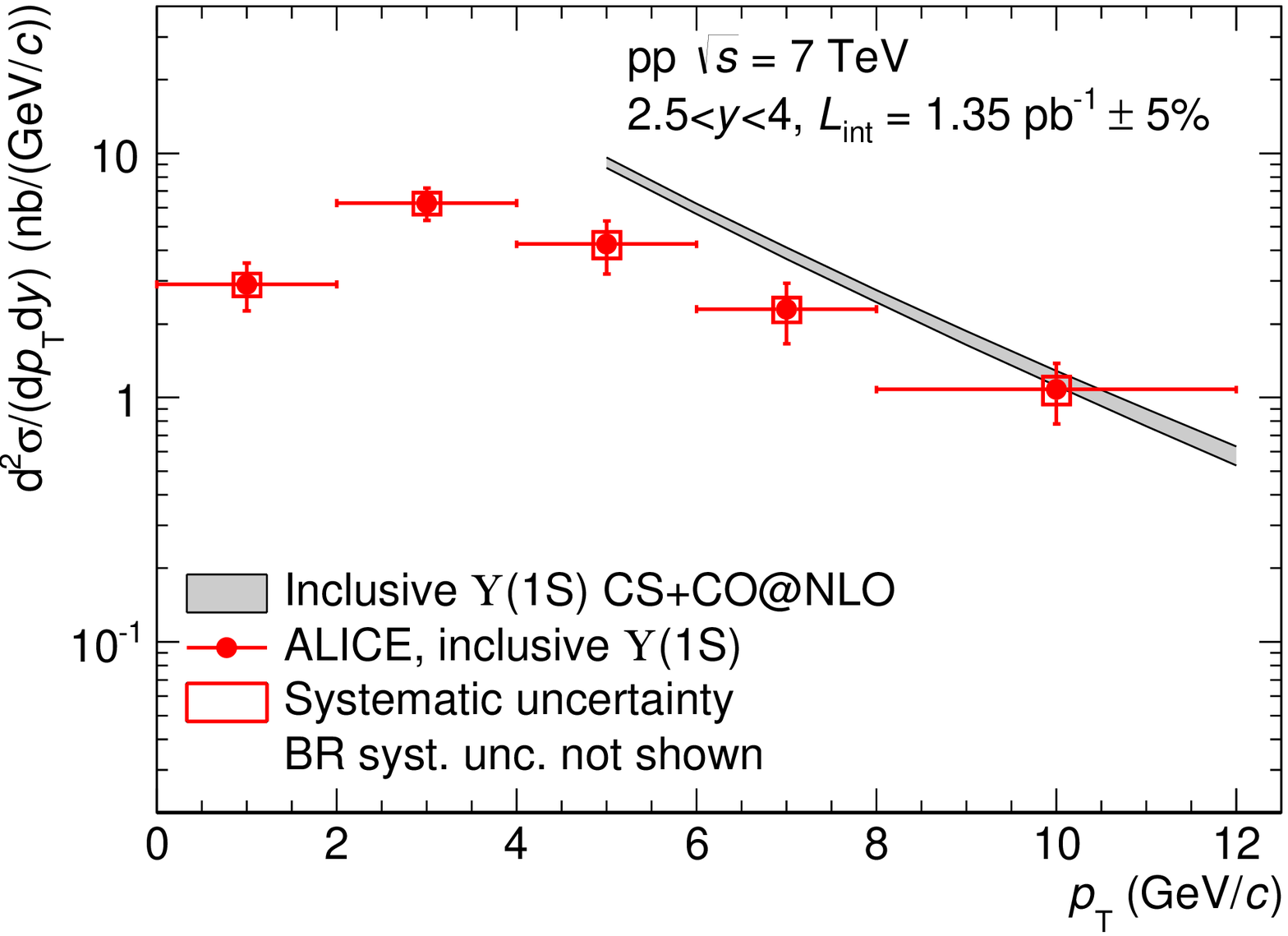}
\caption{(color online). Differential inclusive production cross section of $\rm\Upsilon$(1S) as a function of $\pt$ compared to three scaled CSM calculations of direct $\rm\Upsilon$(1S)~\cite{Lansberg:2012ta} (left) and a NRQCD calculation of inclusive $\rm\Upsilon$(1S)~\cite{Wang:2012is,Wang:2014_ALICE} (right).}
\label{upsilon_theory_pt}
\end{figure}

In Fig.~\ref{upsilon_theory_pt}, the inclusive $\rm\Upsilon$(1S) differential production cross section as a function of $\pt$ is compared to three CSM calculations~\cite{Lansberg:2012ta} (left) and to NRQCD~\cite{Ma:2010jj} (right). 

The CSM calculations are the same as for the $\jpsi$: two complete calculations at LO and NLO respectively and a calculation, called NNLO*, that includes the leading-$\pt$ contributions appearing at NNLO~\cite{Lansberg:2012ta}.
They have been scaled by a factor $1/0.6$ to account for the contributions of $\rm\Upsilon$(2S) ($9$~\%, factor $f^{\rm\Upsilon({\rm 2S})}$, Section~\ref{sec:results}), $\rm\Upsilon$(3S) ($\sim 1$\%~\cite{LHCb:2012aa}) and $\chib$ ($\chib$(1P) $\sim 20$\%~\cite{Aaij:2012se} and $\chib$(2P)$\sim 10$\%~\cite{Affolder:1999wm}) decaying into $\rm\Upsilon$(1S). The comparison between these calculations and the data shows qualitatively the same features as for the $\jpsi$ case: the LO calculation underestimates the data for $\pt>4$~GeV/$c$ and falls too rapidly with increasing $\pt$. The $\pt$ dependence of the NLO calculation is closer to that of the data, but the calculation still underestimates the cross section over the full $\pt$ range. A good agreement is achieved at NNLO, but over a limited $\pt$ range and with large theoretical uncertainties.

The NRQCD calculation is performed by the same group as in Fig.~\ref{psi_theory_pt_com} (right) for the $\jpsi$ and $\psiprime$~\cite{Ma:2010jj}. It includes all the feed-down contributions from $\rm\Upsilon$(2S), $\rm\Upsilon$(3S) and $\chib$. In the limited $\pt$ range of our measurement, the theory overestimates the data. This disagreement becomes smaller for increasing $\pt$ as it is also the case for the LHCb data~\cite{LHCb:2012aa}.

In the CSM, the direct $\rm\Upsilon$(2S) to direct $\rm\Upsilon$(1S) cross section ratio is a constant equal to $0.45$~\cite{Lansberg:2012ta}. In order to compare this value to the measurement quoted in Section~\ref{sec:results}, it must be scaled by the direct-to-inclusive $\rm\Upsilon$(1S) and $\rm\Upsilon$(2S) ratios. For $\rm\Upsilon$(1S), we use a scaling factor of 0.6, as discussed above. For $\rm\Upsilon$(2S), we consider a 5\% contribution from $\rm\Upsilon$(3S)~\cite{LHCb:2012aa} and neglect the contribution from $\chi_b$, which has not been measured to date. We get an upper limit for the $\rm\Upsilon$(2S) direct-to-inclusive ratio of 0.95 and consequently a lower limit for the scaled direct $\rm\Upsilon$(2S)-to-$\rm\Upsilon$(1S) ratio of 0.28. This lower limit is in good agreement with the measurement. We 
note that the measurement is also in good agreement with a NRQCD calculation performed at LO, as described in~\cite{Kisslinger:2013mev}.

\subsection{\label{subsec:theory_y}Differential production cross sections as a function of rapidity}

\begin{figure}[h!]
\includegraphics[width=0.5\linewidth,keepaspectratio]{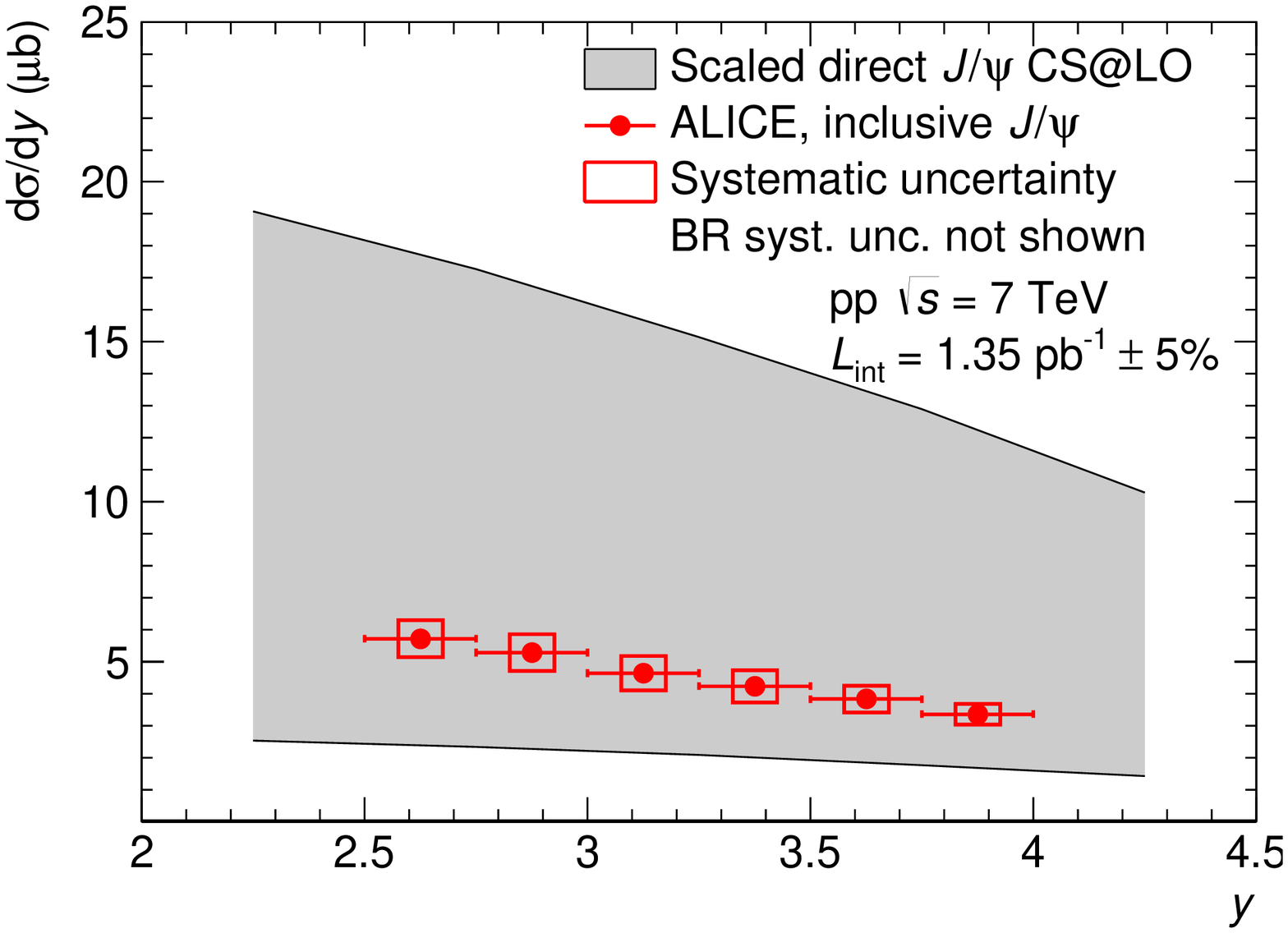}
\includegraphics[width=0.5\linewidth,keepaspectratio]{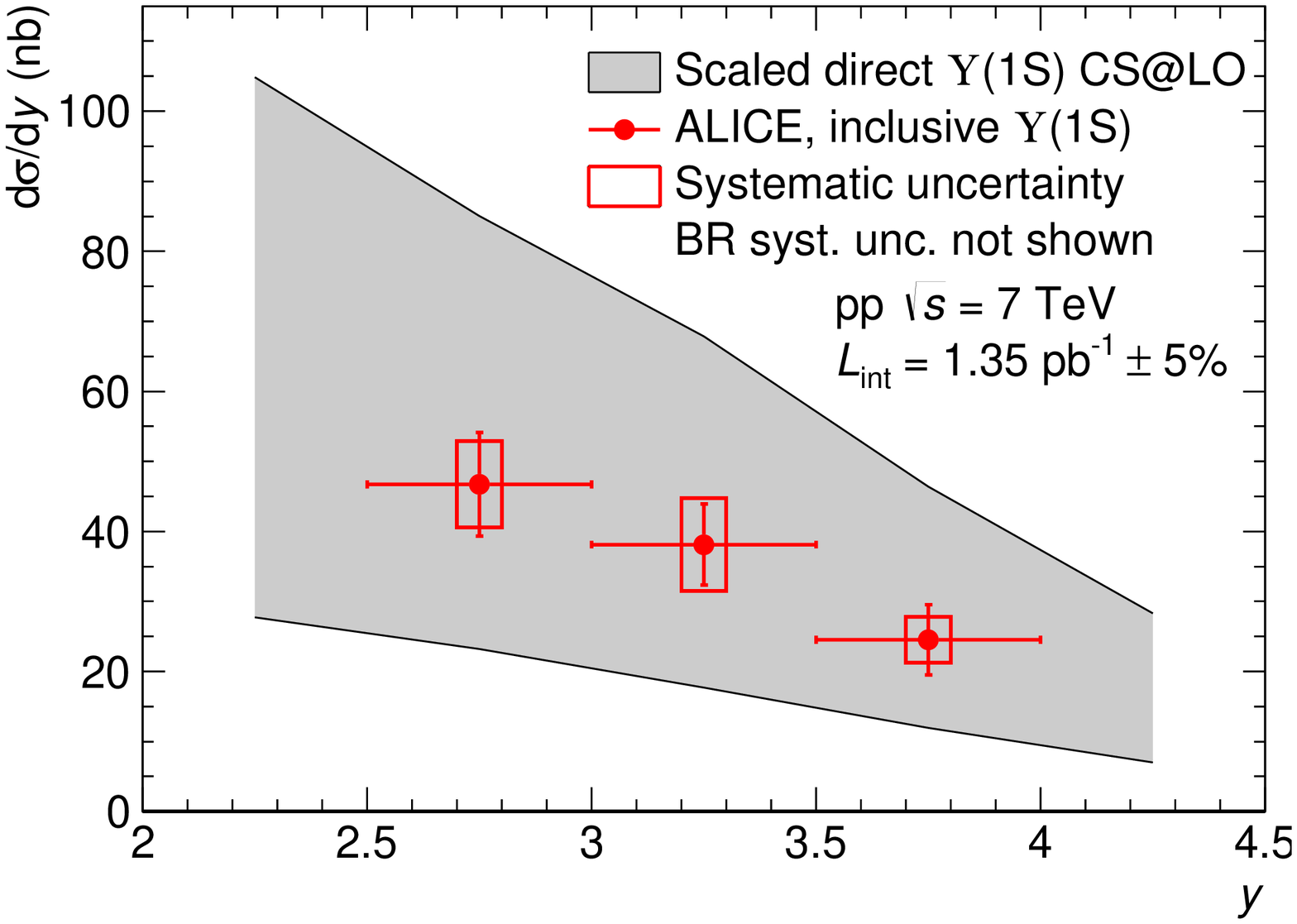}
\caption{(color online). Differential inclusive production cross sections of $\jpsi$ (left) and $\rm\Upsilon$(1S) (right) as a function of $y$ compared to a CSM calculation at LO~\cite{Lansberg:2012ta}.}
\label{upsilon_theory_y}
\end{figure}

Since the LO CSM calculations described in the previous section extend down to zero $\pt$ they can be integrated over $\pt$ and evaluated as a function of the quarkonium rapidity. 
The result is compared to the measured inclusive differential cross sections of $\jpsi$ and $\rm\Upsilon$(1S) in Fig.~\ref{upsilon_theory_y}. 
As for the $\pt$ differential cross sections, the calculations are scaled by the direct-to-inclusive ratios described in the previous section ($1/0.6$ for $\jpsi$ and $\rm\Upsilon$(1S)). 
Extending the calculation down to zero $\pt$ results in large theoretical uncertainties: a factor four to five between the lower and upper bounds. The magnitude of the calculations is in agreement with the measurements. It is also worth noting that these calculations have no free parameters.


\section{\label{sec:conclusion}Conclusion}
In conclusion, the inclusive production cross sections of $\jpsi$, $\psiprime$, $\rm\Upsilon$(1S) and $\rm\Upsilon$(2S) as a function of $\pt$ and $y$ have been measured using the ALICE detector at forward rapidity ($2.5<y<4$) in $\pp$ collisions at a centre of mass energy $\sqrt{s}=7$~TeV. For $\jpsi$, the measurements reported here represent an increase by a factor of about $80$ in terms of luminosity with respect to published ALICE results, whereas they are the first ALICE measurements for the other three quarkonium states. The measured inclusive cross sections, integrated over $\pt$ and $y$ are: $\sigma_{\rm \jpsi}=~6.69\pm0.04\pm0.63$~$\mu$b, $\sigma_\psiprime=1.13\pm0.07\pm0.19$~$\mu$b, $\sigma_{\rm\Upsilon({\rm 1S})}=54.2\pm5.0\pm6.7$~nb and $\sigma_{\rm\Upsilon({\rm 2S})}=18.4\pm3.7\pm2.9$~nb, where the first uncertainty is statistical and the second one is systematic, assuming no quarkonium polarization. Measuring both $\jpsi$ and $\psiprime$ cross sections with the same apparatus and the same data set allows deriving the fraction of inclusive $\jpsi$ that comes from $\psiprime$ decay with reduced systematic uncertainties: $f^{\psiprime}=0.103\pm0.007\pm0.008$. Similarly, the fraction of inclusive $\rm\Upsilon$(1S) that comes from $\rm\Upsilon$(2S) decay is $f^{\rm\Upsilon{\rm(2S)}}=0.090\pm0.027\pm0.005$.

These results are in good agreement with measurements from the LHCb experiment over similar $\pt$ and $y$ ranges.
For $\rm\Upsilon$(1S) and $\rm\Upsilon$(2S) they complement the measurements from CMS at mid-rapidity ($|y|<2.4$).
They are also in good agreement with NRQCD calculations for which the matrix elements have been fitted to data sets from Tevatron, RHIC and LHC, among others.
In the CSM, both LO and NLO calculations underestimate the data at large $\pt$ as it was the case at lower energy.
The addition of the leading-$\pt$ NNLO contributions helps to reduce this disagreement at the price of larger theoretical uncertainties.
LO calculations reproduce qualitatively the data at low $\pt$ and the rapidity dependence of the $\pt$-integrated cross sections.

\clearpage
\section{\label{sec:tables} Integrated and differential quarkonium yields and cross sections}
In the following tables, the systematic uncertainties correspond to the 
quadratic sum of the different sources presented in Section~\ref{systematics} without the contribution from the luminosity and the branching ratios. $\aeff$ corresponds to the acceptance times efficiency factor.

\begin{table}[h]
\centering
\begin{tabular}{ccccc}
\hline
$0<\pt<20$ (GeV/$c$) & $N\pm{\rm stat}$ & $\aeff\pm{\rm stat}$ (\%) & $\sigma\pm{\rm stat}\pm{\rm syst}$ \\
$2.5<y<4$ \\
\hline
$\jpsi$ & $70752\pm371$ &  $13.22\pm0.02$ & $6.69\pm0.04\pm0.53$ $\mu$b\\
\hline
\hline
$0<\pt<12$ (GeV/$c$) & $N\pm{\rm stat}$ & $\aeff\pm{\rm stat}$ (\%) & $\sigma\pm{\rm stat}\pm{\rm syst}$ \\
$2.5<y<4$ \\
\hline
$\psiprime$ & $1987\pm127$ & $16.64\pm0.02$ & $1.13\pm0.07\pm0.12$ $\mu$b\\
$\rm\Upsilon(1S)$ & $380\pm35$ & $20.93\pm0.05$ & $54.23\pm5.01\pm5.98$ nb\\
$\rm\Upsilon(2S)$ & $101\pm20$ & $21.02\pm0.05$ & $18.44\pm3.70\pm2.18$ nb\\
\hline
\end{tabular}

\caption{Integrated raw yields and cross sections of $\jpsi$, $\psiprime$, $\rm\Upsilon(1S)$ and $\rm\Upsilon(2S)$ for $\pp$ collisions at $\sqrt{s}=7$~TeV.}
\label{tab:integrated}
\end{table}

\begin{table}[h]
\centering
\begin{tabular}{cccc}
\hline
$\pt$\ & $N_\jpsi\pm{\rm stat}$ & $\aeff\pm{\rm stat}$ & $d^2\sigma_\jpsi/(d\pt dy)\pm{\rm stat}\pm{\rm syst}$ \\
(GeV/$c$) & &(\%) &($\mu$b/(GeV/$c$))\\
\hline
$[0; 1]$ & $10831\pm161$  & $12.51\pm0.06$  & $0.721\pm0.011\pm0.049$\\
$[1; 2]$ & $17303\pm196$  & $10.67\pm0.04$  & $1.350\pm0.015\pm0.093$\\
$[2; 3]$ & $13859\pm162$  & $10.92\pm0.05$  & $1.057\pm0.012\pm0.068$\\
$[3; 4]$ & $10134\pm133$  & $13.49\pm0.05$  & $0.626\pm0.008\pm0.038$\\
$[4; 5]$ & $7009\pm103$   & $17.20\pm0.06$  & $0.339\pm0.005\pm0.020$\\
$[5; 6]$  & $4398\pm81$   & $21.32\pm0.07$  & $0.172\pm0.003\pm0.011$\\
$[6; 8]$  & $4392\pm80$   & $26.53\pm0.06$  & $0.0689\pm0.0013\pm0.0044$\\
$[8; 10]$ & $1569\pm47$   & $32.75\pm0.06$  & $0.0199\pm0.0006\pm0.0013$\\
$[10; 12]$ & $628\pm31$   & $37.31\pm0.07$  & $0.0070\pm0.0003\pm0.0005$\\
$[12; 14]$ & $287\pm24$   & $40.59\pm0.08$  & $0.0029\pm0.0002\pm0.0002$\\
$[14; 16]$ & $128\pm17$   & $42.95\pm0.08$  & $0.0012\pm0.0002\pm0.0001$\\
$[16; 18]$  & $65\pm11$   & $44.80\pm0.10$  & $0.0006\pm0.0001\pm0.0001$\\
$[18; 20]$  & $33\pm10$   & $46.03\pm0.11$  & $0.0003\pm0.0001\pm0.0001$\\
\hline
\hline
y   & $N_\jpsi\pm{\rm stat}$  & $\aeff\pm{\rm stat}$ (\%)  & $d\sigma_\jpsi/dy\pm{\rm stat}\pm{\rm syst}$ ($\mu$b) \\
\hline
$[2.5; 2.75]$ &  $4660\pm93$  & $4.07\pm0.03$ & $5.72\pm0.11\pm0.60$\\
$[2.75; 3.0]$ & $14768\pm165$ & $13.97\pm0.05$ & $5.28\pm0.06\pm0.59$\\
$[3.0; 3.25]$ & $18559\pm196$ & $19.97\pm0.07$ & $4.64\pm0.05\pm0.55$\\
$[3.25; 3.5]$ & $17241\pm185$ & $20.35\pm0.07$ & $4.23\pm0.05\pm0.50$\\
$[3.5; 3.75]$ & $11727\pm148$ & $15.30\pm0.06$ & $3.83\pm0.05\pm0.43$\\
$[3.75; 4.0]$ &  $3691\pm82$  & $5.49\pm0.03$ & $3.36\pm0.08\pm0.33$\\
\hline
\end{tabular}
\caption{Differential raw yields and cross sections of $\jpsi$ for $\pp$ collisions at $\sqrt{s}=7$~TeV.}
\label{tab:differential}
\end{table}


\begin{table}[h]
\centering
\begin{tabular}{cccc}
\hline
$\pt$\ & $N_\psiprime\pm{\rm stat}$ & $\aeff\pm{\rm stat}$  & $d^2\sigma_{\psiprime)}/(d\pt dy)\pm{\rm stat}\pm{\rm syst}$\\
(GeV/$c$) &&(\%)&($\mu$b/(GeV/$c$))\\
\hline
$[0; 1]$   & $191\pm52$  & $17.63\pm0.07$ & $0.069\pm0.019\pm0.008$\\
$[1; 2]$   & $572\pm73$  & $15.51\pm0.06$ & $0.234\pm0.030\pm0.028$\\
$[2; 3]$   & $350\pm57$  & $14.18\pm0.05$ & $0.156\pm0.025\pm0.017$\\
$[3; 4]$   & $259\pm42$  & $14.87\pm0.06$ & $0.110\pm0.018\pm0.014$\\
$[4; 5]$   & $197\pm30$  & $17.01\pm0.06$ & $0.073\pm0.011\pm0.0090$\\
$[5; 6]$   & $150\pm28$  & $20.15\pm0.07$ & $0.047\pm0.0088\pm0.0059$\\
$[6; 8]$   & $111\pm24$  & $24.81\pm0.05$ & $0.0142\pm0.0031\pm0.0014$\\
$[8; 10]$  &  $69\pm15$  & $30.75\pm0.06$ & $0.0071\pm0.0015\pm0.0007$\\
$[10; 12]$ &  $33\pm11$  & $35.28\pm0.07$ & $0.0030\pm0.0010\pm0.0004$\\
\hline
\hline
y   & $N_\psiprime\pm{\rm stat}$  & $\aeff\pm{\rm stat}$ (\%) & $d\sigma_\psiprime/dy\pm{\rm stat}\pm{\rm{syst}}$ ($\mu$b)\\
\hline
$[2.5; 2.75]$ & $117\pm36$  & $5.63\pm0.03$ &  $0.79\pm0.24\pm0.11$\\
$[2.75; 3.0]$ & $402\pm58$  & $18.10\pm0.06$ & $0.84\pm0.12\pm0.13$\\
$[3.0; 3.25]$ & $538\pm67$  & $25.12\pm0.07$ & $0.81\pm0.10\pm0.12$\\
$[3.25; 3.5]$ & $480\pm63$  & $25.20\pm0.07$ & $0.72\pm0.10\pm0.10$\\
$[3.5; 3.75]$ & $344\pm48$  & $18.67\pm0.06$ & $0.70\pm0.10\pm0.10$\\
$[3.75; 4.0]$ & $ 93\pm26$  & $6.58\pm0.04$ &  $0.54\pm0.15\pm0.07$\\
\hline
\end{tabular}
\caption{Differential raw yields and cross sections of $\psiprime$ for $\pp$ collisions at $\sqrt{s}=7$~TeV.}
\label{tab:j}
\end{table}

\begin{table}[h]
\centering
\begin{tabular}{ccc}
\hline
$\pt$ (GeV/$c$) & $(\frac{d^2\sigma_\psiprime}{d\pt dy})/(\frac{d^2\sigma_\jpsi}{d\pt dy})\pm{\rm stat}\pm{\rm syst}$ \\

\hline
$[0; 1]$ & $0.097\pm0.026\pm0.007$\\
$[1; 2]$ & $0.173\pm0.022\pm0.015$\\
$[2; 3]$ & $0.148\pm0.024\pm0.011$\\
$[3; 4]$ & $0.176\pm0.029\pm0.019$\\
$[4; 5]$ & $0.215\pm0.033\pm0.019$\\
$[5; 6]$ & $0.282\pm0.050\pm0.028$\\
$[6; 8]$ & $0.218\pm0.045\pm0.016$\\
$[8; 10]$ & $0.356\pm0.078\pm0.028$\\
$[10; 12]$ & $0.42\pm0.14\pm0.03$\\
\hline
\hline
$y$  & $(\frac{d\sigma_\psiprime}{dy})/(\frac{d\sigma_\jpsi}{dy})\pm{\rm stat}\pm{\rm syst}$\\
\hline
$[2.5; 2.75]$ & $0.137\pm0.042\pm0.013$\\
$[2.75; 3.0]$ & $0.160\pm0.024\pm0.016$\\
$[3.0; 3.25]$ & $0.175\pm0.022\pm0.014$\\
$[3.25; 3.5]$ & $0.171\pm0.023\pm0.013$\\
$[3.5; 3.75]$ & $0.183\pm0.026\pm0.017$\\
$[3.75; 4.0]$ & $0.160\pm0.046\pm0.017$\\
\hline
\end{tabular}

\vspace{0.3cm}
\caption{Inclusive $\psiprime$-to-$\jpsi$ cross section ratios as a function of
$p_{{\mathrm T}}$ and y for pp collisions at $\sqrt{s}=7$~TeV.}
\label{tab:h}
\end{table}
\begin{table}[h]
\centering
\begin{tabular}{cccc}
\hline
$\pt$ & $N_{\rm\Upsilon({\rm 1S})}\pm{\rm stat}$ & $\aeff\pm{\rm stat}$ & $d^2\sigma_{\rm\Upsilon(1S)}/(d\pt dy)\pm{\rm stat}\pm{\rm syst}$ \\
(GeV/$c$) &&(\%)&(nb/(GeV/$c$))\\
\hline
$[0; 2]$   & $59\pm13  $& $20.21\pm0.18 $& $2.91\pm0.64\pm0.31$\\
$[2; 4]$   & $126\pm19  $& $20.04\pm0.13 $& $6.26\pm0.94\pm0.64$\\
$[4; 6]$   & $86\pm21  $& $20.13\pm0.13 $& $4.25\pm1.04\pm0.53$\\
$[6; 8]$   & $47\pm13  $& $20.38\pm0.16 $& $2.30\pm0.64\pm0.27$\\
$[8; 12]$   & $47\pm13  $& $21.76\pm0.17 $& $1.08\pm0.30\pm0.14$\\
\hline
\hline
$y$ & $N_{\rm\Upsilon({\rm 1S})}\pm{\rm stat}$ & $\aeff\pm{\rm stat}$ (\%) & $d\sigma_{\rm\Upsilon(1S)}/dy\pm{\rm stat}\pm{\rm syst}$ (nb)\\
\hline
$[2.5; 3]$ & $121\pm19  $& $15.47\pm0.10 $& $46.7\pm7.4\pm6.1$\\
$[3; 3.5]$ & $200\pm30  $& $31.34\pm0.13 $& $38.1\pm5.8\pm6.6$\\
$[3.5; 4.0]$ & $67\pm14  $& $16.32\pm0.12 $& $24.5\pm5.0\pm3.3$\\
\hline
\end{tabular}
\caption{Differential raw yields and cross sections of $\rm\Upsilon$(1S) for $\pp$ collisions at $\sqrt{s}=7$~TeV.}
\label{tab:u}
\end{table}

\clearpage               
%
\newenvironment{acknowledgement}{\relax}{\relax}
\begin{acknowledgement}
\section*{Acknowledgements}
We are grateful to M. Butenschoen, K.-T. Chao, R.L. Kisslinger and J.-P. Lansberg for providing us model calculations and for intensive discussions.

The ALICE Collaboration would like to thank all its engineers and technicians for their invaluable contributions to the construction of the experiment and the CERN accelerator teams for the outstanding performance of the LHC complex.
\\
The ALICE Collaboration gratefully acknowledges the resources and support provided by all Grid centres and the Worldwide LHC Computing Grid (WLCG) collaboration.
\\
The ALICE Collaboration acknowledges the following funding agencies for their support in building and
running the ALICE detector:
 \\
State Committee of Science,  World Federation of Scientists (WFS)
and Swiss Fonds Kidagan, Armenia,
 \\
Conselho Nacional de Desenvolvimento Cient\'{\i}fico e Tecnol\'{o}gico (CNPq), Financiadora de Estudos e Projetos (FINEP),
Funda\c{c}\~{a}o de Amparo \`{a} Pesquisa do Estado de S\~{a}o Paulo (FAPESP);
 \\
National Natural Science Foundation of China (NSFC), the Chinese Ministry of Education (CMOE)
and the Ministry of Science and Technology of China (MSTC);
 \\
Ministry of Education and Youth of the Czech Republic;
 \\
Danish Natural Science Research Council, the Carlsberg Foundation and the Danish National Research Foundation;
 \\
The European Research Council under the European Community's Seventh Framework Programme;
 \\
Helsinki Institute of Physics and the Academy of Finland;
 \\
French CNRS-IN2P3, the `Region Pays de Loire', `Region Alsace', `Region Auvergne' and CEA, France;
 \\
German BMBF and the Helmholtz Association;
\\
General Secretariat for Research and Technology, Ministry of
Development, Greece;
\\
Hungarian OTKA and National Office for Research and Technology (NKTH);
 \\
Department of Atomic Energy and Department of Science and Technology of the Government of India;
 \\
Istituto Nazionale di Fisica Nucleare (INFN) and Centro Fermi -
Museo Storico della Fisica e Centro Studi e Ricerche "Enrico
Fermi", Italy;
 \\
MEXT Grant-in-Aid for Specially Promoted Research, Ja\-pan;
 \\
Joint Institute for Nuclear Research, Dubna;
 \\
National Research Foundation of Korea (NRF);
 \\
CONACYT, DGAPA, M\'{e}xico, ALFA-EC and the EPLANET Program
(European Particle Physics Latin American Network)
 \\
Stichting voor Fundamenteel Onderzoek der Materie (FOM) and the Nederlandse Organisatie voor Wetenschappelijk Onderzoek (NWO), Netherlands;
 \\
Research Council of Norway (NFR);
 \\
Polish Ministry of Science and Higher Education;
 \\
National Science Centre, Poland;
 \\
 Ministry of National Education/Institute for Atomic Physics and CNCS-UEFISCDI - Romania;
 \\
Ministry of Education and Science of Russian Federation, Russian
Academy of Sciences, Russian Federal Agency of Atomic Energy,
Russian Federal Agency for Science and Innovations and The Russian
Foundation for Basic Research;
 \\
Ministry of Education of Slovakia;
 \\
Department of Science and Technology, South Africa;
 \\
CIEMAT, EELA, Ministerio de Econom\'{i}a y Competitividad (MINECO) of Spain, Xunta de Galicia (Conseller\'{\i}a de Educaci\'{o}n),
CEA\-DEN, Cubaenerg\'{\i}a, Cuba, and IAEA (International Atomic Energy Agency);
 \\
Swedish Research Council (VR) and Knut $\&$ Alice Wallenberg
Foundation (KAW);
 \\
Ukraine Ministry of Education and Science;
 \\
United Kingdom Science and Technology Facilities Council (STFC);
 \\
The United States Department of Energy, the United States National
Science Foundation, the State of Texas, and the State of Ohio.
\end{acknowledgement}
%
%
\providecommand{\href}[2]{#2}\begingroup\raggedright\endgroup

\newpage

\appendix
\section{The ALICE Collaboration}
\label{app:collab}



\begingroup
\small
\begin{flushleft}
B.~Abelev\Irefn{org69}\And
J.~Adam\Irefn{org37}\And
D.~Adamov\'{a}\Irefn{org77}\And
M.M.~Aggarwal\Irefn{org81}\And
M.~Agnello\Irefn{org104}\textsuperscript{,}\Irefn{org87}\And
A.~Agostinelli\Irefn{org26}\And
N.~Agrawal\Irefn{org44}\And
Z.~Ahammed\Irefn{org123}\And
N.~Ahmad\Irefn{org18}\And
A.~Ahmad~Masoodi\Irefn{org18}
% ...
\renewcommand\labelenumi{\textsuperscript{\theenumi}~}

\section*{Affiliation notes}
\renewcommand\theenumi{\roman{enumi}}
\begin{Authlist}
\item \Adef{0}Deceased
\item \Adef{idp1106880}{Also at: St. Petersburg State Polytechnical University}
\item \Adef{idp2956864}{Also at: Department of Applied Physics, Aligarh Muslim University, Aligarh, India}
\item \Adef{idp3641296}{Also at: M.V. Lomonosov Moscow State University, D.V. Skobeltsyn Institute of Nuclear Physics, Moscow, Russia}
\item \Adef{idp3884896}{Also at: University of Belgrade, Faculty of Physics and "Vin\v{c}a" Institute of Nuclear Sciences, Belgrade, Serbia}
\item \Adef{idp4165744}{Permanent Address: Permanent Address: Konkuk University, Seoul, Korea}
\item \Adef{idp4711504}{Also at: Institute of Theoretical Physics, University of Wroclaw, Wroclaw, Poland}
\item \Adef{idp5597984}{Also at: University of Kansas, Lawrence, KS, United States}
\end{Authlist}

\section*{Collaboration Institutes}
\renewcommand\theenumi{\arabic{enumi}~}
\begin{Authlist}

\item \Idef{org1}A.I. Alikhanyan National Science Laboratory (Yerevan Physics Institute) Foundation, Yerevan, Armenia
\item \Idef{org2}Benem\'{e}rita Universidad Aut\'{o}noma de Puebla, Puebla, Mexico
\item \Idef{org3}Bogolyubov Institute for Theoretical Physics, Kiev, Ukraine
\item \Idef{org4}Bose Institute, Department of Physics and Centre for Astroparticle Physics and Space Science (CAPSS), Kolkata, India
\item \Idef{org5}Budker Institute for Nuclear Physics, Novosibirsk, Russia
\item \Idef{org6}California Polytechnic State University, San Luis Obispo, CA, United States
\item \Idef{org7}Central China Normal University, Wuhan, China
\item \Idef{org8}Centre de Calcul de l'IN2P3, Villeurbanne, France
\item \Idef{org9}Centro de Aplicaciones Tecnol\'{o}gicas y Desarrollo Nuclear (CEADEN), Havana, Cuba
\item \Idef{org10}Centro de Investigaciones Energ\'{e}ticas Medioambientales y Tecnol\'{o}gicas (CIEMAT), Madrid, Spain
\item \Idef{org11}Centro de Investigaci\'{o}n y de Estudios Avanzados (CINVESTAV), Mexico City and M\'{e}rida, Mexico
\item \Idef{org12}Centro Fermi - Museo Storico della Fisica e Centro Studi e Ricerche ``Enrico Fermi'', Rome, Italy
\item \Idef{org13}Chicago State University, Chicago, USA
\item \Idef{org14}Commissariat \`{a} l'Energie Atomique, IRFU, Saclay, France
\item \Idef{org15}COMSATS Institute of Information Technology (CIIT), Islamabad, Pakistan
\item \Idef{org16}Departamento de F\'{\i}sica de Part\'{\i}culas and IGFAE, Universidad de Santiago de Compostela, Santiago de Compostela, Spain
\item \Idef{org17}Department of Physics and Technology, University of Bergen, Bergen, Norway
\item \Idef{org18}Department of Physics, Aligarh Muslim University, Aligarh, India
\item \Idef{org19}Department of Physics, Ohio State University, Columbus, OH, United States
\item \Idef{org20}Department of Physics, Sejong University, Seoul, South Korea
\item \Idef{org21}Department of Physics, University of Oslo, Oslo, Norway
\item \Idef{org22}Dipartimento di Fisica dell'Universit\`{a} 'La Sapienza' and Sezione INFN Rome, Italy
\item \Idef{org23}Dipartimento di Fisica dell'Universit\`{a} and Sezione INFN, Cagliari, Italy
\item \Idef{org24}Dipartimento di Fisica dell'Universit\`{a} and Sezione INFN, Trieste, Italy
\item \Idef{org25}Dipartimento di Fisica dell'Universit\`{a} and Sezione INFN, Turin, Italy
\item \Idef{org26}Dipartimento di Fisica e Astronomia dell'Universit\`{a} and Sezione INFN, Bologna, Italy
\item \Idef{org27}Dipartimento di Fisica e Astronomia dell'Universit\`{a} and Sezione INFN, Catania, Italy
\item \Idef{org28}Dipartimento di Fisica e Astronomia dell'Universit\`{a} and Sezione INFN, Padova, Italy
\item \Idef{org29}Dipartimento di Fisica `E.R.~Caianiello' dell'Universit\`{a} and Gruppo Collegato INFN, Salerno, Italy
\item \Idef{org30}Dipartimento di Scienze e Innovazione Tecnologica dell'Universit\`{a} del  Piemonte Orientale and Gruppo Collegato INFN, Alessandria, Italy
\item \Idef{org31}Dipartimento Interateneo di Fisica `M.~Merlin' and Sezione INFN, Bari, Italy
\item \Idef{org32}Division of Experimental High Energy Physics, University of Lund, Lund, Sweden
\item \Idef{org33}Eberhard Karls Universit\"{a}t T\"{u}bingen, T\"{u}bingen, Germany
\item \Idef{org34}European Organization for Nuclear Research (CERN), Geneva, Switzerland
\item \Idef{org35}Faculty of Engineering, Bergen University College, Bergen, Norway
\item \Idef{org36}Faculty of Mathematics, Physics and Informatics, Comenius University, Bratislava, Slovakia
\item \Idef{org37}Faculty of Nuclear Sciences and Physical Engineering, Czech Technical University in Prague, Prague, Czech Republic
\item \Idef{org38}Faculty of Science, P.J.~\v{S}af\'{a}rik University, Ko\v{s}ice, Slovakia
\item \Idef{org39}Frankfurt Institute for Advanced Studies, Johann Wolfgang Goethe-Universit\"{a}t Frankfurt, Frankfurt, Germany
\item \Idef{org40}Gangneung-Wonju National University, Gangneung, South Korea
\item \Idef{org41}Gauhati University, Department of Physics, Guwahati, India
\item \Idef{org42}Helsinki Institute of Physics (HIP), Helsinki, Finland
\item \Idef{org43}Hiroshima University, Hiroshima, Japan
\item \Idef{org44}Indian Institute of Technology Bombay (IIT), Mumbai, India
\item \Idef{org45}Indian Institute of Technology Indore, Indore (IITI), India
\item \Idef{org46}Institut de Physique Nucl\'eaire d'Orsay (IPNO), Universit\'e Paris-Sud, CNRS-IN2P3, Orsay, France
\item \Idef{org47}Institut f\"{u}r Informatik, Johann Wolfgang Goethe-Universit\"{a}t Frankfurt, Frankfurt, Germany
\item \Idef{org48}Institut f\"{u}r Kernphysik, Johann Wolfgang Goethe-Universit\"{a}t Frankfurt, Frankfurt, Germany
\item \Idef{org49}Institut f\"{u}r Kernphysik, Westf\"{a}lische Wilhelms-Universit\"{a}t M\"{u}nster, M\"{u}nster, Germany
\item \Idef{org50}Institut Pluridisciplinaire Hubert Curien (IPHC), Universit\'{e} de Strasbourg, CNRS-IN2P3, Strasbourg, France
\item \Idef{org51}Institute for Nuclear Research, Academy of Sciences, Moscow, Russia
\item \Idef{org52}Institute for Subatomic Physics of Utrecht University, Utrecht, Netherlands
\item \Idef{org53}Institute for Theoretical and Experimental Physics, Moscow, Russia
\item \Idef{org54}Institute of Experimental Physics, Slovak Academy of Sciences, Ko\v{s}ice, Slovakia
\item \Idef{org55}Institute of Physics, Academy of Sciences of the Czech Republic, Prague, Czech Republic
\item \Idef{org56}Institute of Physics, Bhubaneswar, India
\item \Idef{org57}Institute of Space Science (ISS), Bucharest, Romania
\item \Idef{org58}Instituto de Ciencias Nucleares, Universidad Nacional Aut\'{o}noma de M\'{e}xico, Mexico City, Mexico
\item \Idef{org59}Instituto de F\'{\i}sica, Universidad Nacional Aut\'{o}noma de M\'{e}xico, Mexico City, Mexico
\item \Idef{org60}iThemba LABS, National Research Foundation, Somerset West, South Africa
\item \Idef{org61}Joint Institute for Nuclear Research (JINR), Dubna, Russia
\item \Idef{org62}Korea Institute of Science and Technology Information, Daejeon, South Korea
\item \Idef{org63}KTO Karatay University, Konya, Turkey
\item \Idef{org64}Laboratoire de Physique Corpusculaire (LPC), Clermont Universit\'{e}, Universit\'{e} Blaise Pascal, CNRS--IN2P3, Clermont-Ferrand, France
\item \Idef{org65}Laboratoire de Physique Subatomique et de Cosmologie, Universit\'{e} Grenoble-Alpes, CNRS-IN2P3, Grenoble, France
\item \Idef{org66}Laboratori Nazionali di Frascati, INFN, Frascati, Italy
\item \Idef{org67}Laboratori Nazionali di Legnaro, INFN, Legnaro, Italy
\item \Idef{org68}Lawrence Berkeley National Laboratory, Berkeley, CA, United States
\item \Idef{org69}Lawrence Livermore National Laboratory, Livermore, CA, United States
\item \Idef{org70}Moscow Engineering Physics Institute, Moscow, Russia
\item \Idef{org71}National Centre for Nuclear Studies, Warsaw, Poland
\item \Idef{org72}National Institute for Physics and Nuclear Engineering, Bucharest, Romania
\item \Idef{org73}National Institute of Science Education and Research, Bhubaneswar, India
\item \Idef{org74}Niels Bohr Institute, University of Copenhagen, Copenhagen, Denmark
\item \Idef{org75}Nikhef, National Institute for Subatomic Physics, Amsterdam, Netherlands
\item \Idef{org76}Nuclear Physics Group, STFC Daresbury Laboratory, Daresbury, United Kingdom
\item \Idef{org77}Nuclear Physics Institute, Academy of Sciences of the Czech Republic, \v{R}e\v{z} u Prahy, Czech Republic
\item \Idef{org78}Oak Ridge National Laboratory, Oak Ridge, TN, United States
\item \Idef{org79}Petersburg Nuclear Physics Institute, Gatchina, Russia
\item \Idef{org80}Physics Department, Creighton University, Omaha, NE, United States
\item \Idef{org81}Physics Department, Panjab University, Chandigarh, India
\item \Idef{org82}Physics Department, University of Athens, Athens, Greece
\item \Idef{org83}Physics Department, University of Cape Town, Cape Town, South Africa
\item \Idef{org84}Physics Department, University of Jammu, Jammu, India
\item \Idef{org85}Physics Department, University of Rajasthan, Jaipur, India
\item \Idef{org86}Physikalisches Institut, Ruprecht-Karls-Universit\"{a}t Heidelberg, Heidelberg, Germany
\item \Idef{org87}Politecnico di Torino, Turin, Italy
\item \Idef{org88}Purdue University, West Lafayette, IN, United States
\item \Idef{org89}Pusan National University, Pusan, South Korea
\item \Idef{org90}Research Division and ExtreMe Matter Institute EMMI, GSI Helmholtzzentrum f\"ur Schwerionenforschung, Darmstadt, Germany
\item \Idef{org91}Rudjer Bo\v{s}kovi\'{c} Institute, Zagreb, Croatia
\item \Idef{org92}Russian Federal Nuclear Center (VNIIEF), Sarov, Russia
\item \Idef{org93}Russian Research Centre Kurchatov Institute, Moscow, Russia
\item \Idef{org94}Saha Institute of Nuclear Physics, Kolkata, India
\item \Idef{org95}School of Physics and Astronomy, University of Birmingham, Birmingham, United Kingdom
\item \Idef{org96}Secci\'{o}n F\'{\i}sica, Departamento de Ciencias, Pontificia Universidad Cat\'{o}lica del Per\'{u}, Lima, Peru
\item \Idef{org97}Sezione INFN, Bari, Italy
\item \Idef{org98}Sezione INFN, Bologna, Italy
\item \Idef{org99}Sezione INFN, Cagliari, Italy
\item \Idef{org100}Sezione INFN, Catania, Italy
\item \Idef{org101}Sezione INFN, Padova, Italy
\item \Idef{org102}Sezione INFN, Rome, Italy
\item \Idef{org103}Sezione INFN, Trieste, Italy
\item \Idef{org104}Sezione INFN, Turin, Italy
\item \Idef{org105}SSC IHEP of NRC Kurchatov institute, Protvino, Russia
\item \Idef{org106}SUBATECH, Ecole des Mines de Nantes, Universit\'{e} de Nantes, CNRS-IN2P3, Nantes, France
\item \Idef{org107}Suranaree University of Technology, Nakhon Ratchasima, Thailand
\item \Idef{org108}Technical University of Split FESB, Split, Croatia
\item \Idef{org109}The Henryk Niewodniczanski Institute of Nuclear Physics, Polish Academy of Sciences, Cracow, Poland
\item \Idef{org110}The University of Texas at Austin, Physics Department, Austin, TX, USA
\item \Idef{org111}Universidad Aut\'{o}noma de Sinaloa, Culiac\'{a}n, Mexico
\item \Idef{org112}Universidade de S\~{a}o Paulo (USP), S\~{a}o Paulo, Brazil
\item \Idef{org113}Universidade Estadual de Campinas (UNICAMP), Campinas, Brazil
\item \Idef{org114}University of Houston, Houston, TX, United States
\item \Idef{org115}University of Jyv\"{a}skyl\"{a}, Jyv\"{a}skyl\"{a}, Finland
\item \Idef{org116}University of Liverpool, Liverpool, United Kingdom
\item \Idef{org117}University of Tennessee, Knoxville, TN, United States
\item \Idef{org118}University of Tokyo, Tokyo, Japan
\item \Idef{org119}University of Tsukuba, Tsukuba, Japan
\item \Idef{org120}University of Zagreb, Zagreb, Croatia
\item \Idef{org121}Universit\'{e} de Lyon, Universit\'{e} Lyon 1, CNRS/IN2P3, IPN-Lyon, Villeurbanne, France
\item \Idef{org122}V.~Fock Institute for Physics, St. Petersburg State University, St. Petersburg, Russia
\item \Idef{org123}Variable Energy Cyclotron Centre, Kolkata, India
\item \Idef{org124}Vestfold University College, Tonsberg, Norway
\item \Idef{org125}Warsaw University of Technology, Warsaw, Poland
\item \Idef{org126}Wayne State University, Detroit, MI, United States
\item \Idef{org127}Wigner Research Centre for Physics, Hungarian Academy of Sciences, Budapest, Hungary
\item \Idef{org128}Yale University, New Haven, CT, United States
\item \Idef{org129}Yonsei University, Seoul, South Korea
\item \Idef{org130}Zentrum f\"{u}r Technologietransfer und Telekommunikation (ZTT), Fachhochschule Worms, Worms, Germany
\end{Authlist}
\endgroup


\begin{thebibliography}{10}

\bibitem{Cacciari:1998it}
M.~Cacciari, M.~Greco, and P.~Nason, {\em J. High Energy Phys.} {\bfseries
  9805} (1998) 007,
\href{http://arxiv.org/abs/9803400}{{\ttfamily arXiv:9803400 [hep-ph]}}.

\bibitem{Cacciari:2001td}
M.~Cacciari, S.~Frixione, and P.~Nason, {\em J. High Energy Phys.} {\bfseries
  0103} (2001) 006,
\href{http://arxiv.org/abs/0102134}{{\ttfamily arXiv:0102134 [hep-ph]}}.

\bibitem{Cacciari:2012ny}
M.~Cacciari {\em et~al.},
  \href{http://dx.doi.org/10.1007/JHEP10(2012)137}{{\em J. High Energy Phys.}
  {\bfseries 1210} (2012) 137},
\href{http://arxiv.org/abs/1205.6344}{{\ttfamily arXiv:1205.6344 [hep-ph]}}.

\bibitem{Brambilla:2010cs}
N.~Brambilla {\em et~al.},
  \href{http://dx.doi.org/10.1140/epjc/s10052-010-1534-9}{{\em Eur. Phys. J. C}
  {\bfseries 71} (2011) 1534},
\href{http://arxiv.org/abs/1010.5827}{{\ttfamily arXiv:1010.5827 [hep-ph]}}.

\bibitem{Einhorn:1975ua}
 M.B. Einhorn and S.D. Ellis,
{{\em Phys. Rev. D} {\bfseries 12} (1975) 2007}.

\bibitem{Carlson:1976cd}
 C.E. Carlson and R. Suaya,
{{\em Phys. Rev. D} {\bfseries 14} (1976) 3115}.

\bibitem{Baier:1981uk}
R.~Baier and R.~Rückl,
  \href{http://dx.doi.org/http://dx.doi.org/10.1016/0370-2693(81)90636-5}{{\em
  Phys. Lett. B} {\bfseries 102} no.~5, (1981) 364}.

\bibitem{Abe:1997jz}
F.~Abe {\em et~al.} ({CDF Collaboration}),
\href{http://dx.doi.org/10.1103/PhysRevLett.79.572}{{\em Phys. Rev. Lett.}
  {\bfseries 79} (1997) 572}.

\bibitem{PhysRevLett.98.252002}
J.~M. Campbell, F.~Maltoni, and F.~Tramontano,
  \href{http://dx.doi.org/10.1103/PhysRevLett.98.252002}{{\em Phys. Rev. Lett.}
  {\bfseries 98} (2007) 252002},
\href{http://arxiv.org/abs/hep-ph/0703113}{{\ttfamily arXiv:hep-ph/0703113
  [HEP-PH]}}.

\bibitem{Artoisenet:2008fc}
P.~Artoisenet {\em et~al.},
  \href{http://dx.doi.org/10.1103/PhysRevLett.101.152001}{{\em Phys. Rev.
  Lett.} {\bfseries 101} (2008) 152001},
\href{http://arxiv.org/abs/0806.3282}{{\ttfamily arXiv:0806.3282 [hep-ph]}}.

\bibitem{Lansberg:2008gk}
J.-P. Lansberg, \href{http://dx.doi.org/10.1140/epjc/s10052-008-0826-9}{{\em
  Eur. Phys. J. C} {\bfseries 61} (2009) 693},
\href{http://arxiv.org/abs/0811.4005}{{\ttfamily arXiv:0811.4005 [hep-ph]}}.

\bibitem{Braaten:1994xb}
E.~Braaten {\em et~al.},
  \href{http://dx.doi.org/10.1016/0370-2693(94)90182-1}{{\em Phys. Lett. B}
  {\bfseries 333} (1994) 548},
\href{http://arxiv.org/abs/9405407}{{\ttfamily arXiv:9405407 [hep-ph]}}.

\bibitem{Artoisenet:2007xi}
P.~Artoisenet, J.-P. Lansberg, and F.~Maltoni,
  \href{http://dx.doi.org/10.1016/j.physletb.2007.04.031}{{\em Phys. Lett. B}
  {\bfseries 653} (2007) 60},
\href{http://arxiv.org/abs/0703129}{{\ttfamily arXiv:0703129 [hep-ph]}}.

\bibitem{Haberzettl:2007kj}
   H.~Haberzettl and J.~P.~Lansberg,
   {{\em Phys. Rev. Lett.} {\bfseries 100} (2008) 032006},
  {\ttfamily arXiv:0709.3471 [hep-ph]}.

\bibitem{Halzen:1977rs}
F.~Halzen,
\href{http://dx.doi.org/10.1016/0370-2693(77)90144-7}{{\em Phys. Lett. B}
  {\bfseries 69} (1977) 105}.

\bibitem{Fritzsch:1977ay}
H.~Fritzsch,
\href{http://dx.doi.org/10.1016/0370-2693(77)90108-3}{{\em Phys. Lett. B}
  {\bfseries 67} (1977) 217}.

\bibitem{Amundson:1996qr}
J.~Amundson {\em et~al.},
  \href{http://dx.doi.org/10.1016/S0370-2693(96)01417-7}{{\em Phys. Lett. B}
  {\bfseries 390} (1997) 323},
\href{http://arxiv.org/abs/9605295}{{\ttfamily arXiv:9605295 [hep-ph]}}.

\bibitem{Bodwin:1994jh}
G.~T. Bodwin, E.~Braaten, and G.~P. Lepage,
  \href{http://dx.doi.org/10.1103/PhysRevD.55.5853,
  10.1103/PhysRevD.51.1125}{{\em Phys. Rev. D} {\bfseries 51} (1995) 1125},
\href{http://arxiv.org/abs/9407339}{{\ttfamily arXiv:9407339 [hep-ph]}}.

\bibitem{Ma:2010jj}
Y.~Q. Ma, K.~Wang, and K.~T. Chao,
  \href{http://dx.doi.org/10.1103/PhysRevD.84.114001}{{\em Phys. Rev. D}
  {\bfseries 84} (2011) 114001},
\href{http://arxiv.org/abs/1012.1030}{{\ttfamily arXiv:1012.1030 [hep-ph]}}.

\bibitem{Butenschoen:2010rq}
M.~Butenschoen and B.~A. Kniehl,
  \href{http://dx.doi.org/10.1103/PhysRevLett.106.022003}{{\em Phys. Rev.
  Lett.} {\bfseries 106} (2011) 022003},
\href{http://arxiv.org/abs/1009.5662}{{\ttfamily arXiv:1009.5662 [hep-ph]}}.

\bibitem{Xu:2012am}
G.~Xu {\em et~al.}, \href{http://dx.doi.org/10.1103/PhysRevD.86.094017}{{\em
  Phys. Rev. D} {\bfseries 86} (2012) 094017},
\href{http://arxiv.org/abs/1203.0207}{{\ttfamily arXiv:1203.0207 [hep-ph]}}.

\bibitem{Abulencia:2007us}
A.~Abulencia {\em et~al.} ({CDF Collaboration}),
  \href{http://dx.doi.org/10.1103/PhysRevLett.99.132001}{{\em Phys. Rev. Lett.}
  {\bfseries 99} (2007) 132001},
\href{http://arxiv.org/abs/0704.0638}{{\ttfamily arXiv:0704.0638 [hep-ex]}}.

\bibitem{alice:2012pol}
B.~Abelev {\em et~al.} ({ALICE Collaboration}),
  \href{http://dx.doi.org/10.1103/PhysRevLett.108.082001}{{\em Phys. Rev.
  Lett.} {\bfseries 108} (2012) 082001},
\href{http://arxiv.org/abs/1111.1630}{{\ttfamily arXiv:1111.1630 [hep-ex]}}.

\bibitem{Aaij:2013nlm}
R.~Aaij {\em et~al.} ({LHCb Collaboration}),
  \href{http://dx.doi.org/10.1140/epjc/s10052-013-2631-3}{{\em Eur. Phys. J. C}
  {\bfseries 73} (2013) 2631},
\href{http://arxiv.org/abs/1307.6379}{{\ttfamily arXiv:1307.6379 [hep-ex]}}.

\bibitem{Chatrchyan:2013cla}
S.~Chatrchyan {\em et~al.} ({CMS Collaboration}),
\href{http://arxiv.org/abs/1307.6070}{{\ttfamily arXiv:1307.6070 [hep-ex]}}.

\bibitem{Chatrchyan:2012woa}
S.~Chatrchyan {\em et~al.} ({CMS Collaboration}),
  \href{http://dx.doi.org/10.1103/PhysRevLett.110.081802}{{\em Phys. Rev.
  Lett.} {\bfseries 110} (2013) 081802},
\href{http://arxiv.org/abs/1209.2922}{{\ttfamily arXiv:1209.2922 [hep-ex]}}.

\bibitem{Abe:1995an}
F.~Abe {\em et~al.} ({CDF Collaboration}),
\href{http://dx.doi.org/10.1103/PhysRevLett.75.4358}{{\em Phys. Rev. Lett.}
  {\bfseries 75} (1995) 4358}.

\bibitem{LHCb:2012aa}
R.~Aaij {\em et~al.} ({LHCb Collaboration}),
  \href{http://dx.doi.org/10.1140/epjc/s10052-012-2025-y}{{\em Eur. Phys. J. C}
  {\bfseries 72} (2012) 2025},
\href{http://arxiv.org/abs/1202.6579}{{\ttfamily arXiv:1202.6579 [hep-ex]}}.

\bibitem{Aamodt:2011gj}
K.~Aamodt {\em et~al.} ({ALICE Collaboration}),
  \href{http://dx.doi.org/10.1016/j.physletb.2011.09.054,
  10.1016/j.physletb.2012.10.060}{{\em Phys. Lett. B} {\bfseries 704} (2011)
  442}, \href{http://arxiv.org/abs/1105.0380}{{\ttfamily arXiv:1105.0380
  [hep-ex]}}.

\bibitem{Aamodt:2011gjE}
K.~Aamodt {\em et~al.} ({ALICE Collaboration}),
  \href{http://dx.doi.org/http://dx.doi.org/10.1016/j.physletb.2012.10.060}{{\em Phys. Lett. B} {\bfseries 718} no.~2, (2012) 692}.

\bibitem{ALICE}
K.~Aamodt {\em et~al.} ({ALICE Collaboration}),
\href{http://dx.doi.org/10.1088/1748-0221/3/08/S08002}{{\em J. Instrum.}
  {\bfseries 3} (2008) S08002}.

\bibitem{ALICE_ITS}
K.~Aamodt {\em et~al.} ({ALICE Collaboration}),
  \href{http://dx.doi.org/10.1088/1748-0221/5/03/P03003}{{\em J. Instrum.}
  {\bfseries 5} (2010) 3003},
  \href{http://arxiv.org/abs/1001.0502}{{\ttfamily arXiv:1001.0502
  [physics.ins-det]}}.

\bibitem{Abbas:2013taa}
E.~Abbas {\em et~al.} ({ALICE Collaboration}),
\href{http://arxiv.org/abs/1306.3130}{{\ttfamily arXiv:1306.3130 [nucl-ex]}}.

\bibitem{Bondila:2005xy}
M.~Bondila {\em et~al.},
\href{http://dx.doi.org/10.1109/TNS.2005.856900}{{\em IEEE Trans. Nucl. Sci.}
  {\bfseries 52} (2005) 1705}.

\bibitem{mtrBossu2012}
F.~Boss\`u, M.~Gagliardi, and M.~Marchisone ({ALICE Collaboration}), {\em J.
  Instrum.} {\bfseries 7} no.~12, (2012) T12002.

\bibitem{mtr2006}
R.~Arnaldi {\em et~al.},
\href{http://dx.doi.org/10.1016/j.nuclphysbps.2006.07.012}{{\em Nucl. Phys.
  Proc. Suppl.} {\bfseries 158} (2006) 21--24}.

\bibitem{vanderMeer:1968zz}
S.~van~der Meer, {ISR-PO/68-31, KEK68-64}.
 
\bibitem{Abelev:2012sea}
B.~Abelev {\em et~al.} ({ALICE Collaboration}),
  \href{http://dx.doi.org/10.1140/epjc/s10052-013-2456-0}{{\em Eur. Phys. J. C}
  {\bfseries 73} (2013) 2456},
\href{http://arxiv.org/abs/1208.4968}{{\ttfamily arXiv:1208.4968 [hep-ex]}}.

\bibitem{PhysRevD.86.010001}
J.~Beringer {\em et~al.} ({{Particle Data Group}}),
  \href{http://dx.doi.org/10.1103/PhysRevD.86.010001}{{\em Phys. Rev. D}
  {\bfseries 86} (2012) 010001}.

\bibitem{Gaiser:1982yw}
J.~Gaiser {\em et~al.},
\href{http://dx.doi.org/10.1103/PhysRevD.34.711}{{\em Phys. Rev. D} {\bfseries
  34} (1986) 711}.

\bibitem{Shahoyan:2001sd}
R.~Shahoyan, {$\jpsi$ and $\psiprime$ production in 450 GeV pA interactions and
  its dependence on the rapidity and $x_F$}.
\newblock PhD thesis,
2001.
\newblock

\bibitem{Aaij:2011jh}
R.~Aaij {\em et~al.} ({LHCb Collaboration}),
  \href{http://dx.doi.org/10.1140/epjc/s10052-011-1645-y}{{\em Eur. Phys. J. C}
  {\bfseries 71} (2011) 1645},
\href{http://arxiv.org/abs/1103.0423}{{\ttfamily arXiv:1103.0423 [hep-ex]}}.

\bibitem{Aaij:2012ag}
R.~Aaij {\em et~al.} ({LHCb Collaboration}),
  \href{http://dx.doi.org/10.1140/epjc/s10052-012-2100-4}{{\em Eur. Phys. J. C}
  {\bfseries 72} (2012) 2100},
\href{http://arxiv.org/abs/1204.1258}{{\ttfamily arXiv:1204.1258 [hep-ex]}}.

\bibitem{GEANT3}
R.~Brun {\em et~al.}, {\em CERN Program Library Long Write-up W5013} (1994) .

\bibitem{Khachatryan:2010zg}
V.~Khachatryan {\em et~al.} ({CMS Collaboration}),
  \href{http://dx.doi.org/10.1103/PhysRevD.83.112004}{{\em Phys. Rev. D}
  {\bfseries 83} (2011) 112004},
\href{http://arxiv.org/abs/1012.5545}{{\ttfamily arXiv:1012.5545 [hep-ex]}}.

\bibitem{Chatrchyan:2013yna}
S.~Chatrchyan {\em et~al.} ({CMS Collaboration}),
{{\em Phys. Lett. B} {\bfseries 727} (2013) 101},
\href{http://arxiv.org/abs/1303.5900}{{\ttfamily arXiv:1303.5900 [hep-ex]}}.

\bibitem{Lansberg:2011hi}
J.-P. Lansberg, \href{http://dx.doi.org/10.1088/0954-3899/38/12/124110}{{\em J.
  Phys. G} {\bfseries 38} (2011) 124110},
\href{http://arxiv.org/abs/1107.0292}{{\ttfamily arXiv:1107.0292 [hep-ph]}}.

\bibitem{LHCb:2012af}
R.~Aaij {\em et~al.} ({LHCb Collaboration}),
  \href{http://dx.doi.org/10.1016/j.physletb.2012.10.068}{{\em Phys. Lett. B}
  {\bfseries 718} (2012) 431},
\href{http://arxiv.org/abs/1204.1462}{{\ttfamily arXiv:1204.1462 [hep-ex]}}.

\bibitem{Butenschoen:2011yh}
M.~Butenschoen and B.~A. Kniehl,
  \href{http://dx.doi.org/10.1103/PhysRevD.84.051501}{{\em Phys. Rev. D}
  {\bfseries 84} (2011) 051501},
\href{http://arxiv.org/abs/1105.0820}{{\ttfamily arXiv:1105.0820 [hep-ph]}}.

\bibitem{Eichten:1995ch}
E.~J. Eichten and C.~Quigg,
  \href{http://dx.doi.org/10.1103/PhysRevD.52.1726}{{\em Phys. Rev. D}
  {\bfseries 52} (1995) 1726},
\href{http://arxiv.org/abs/9503356}{{\ttfamily arXiv:9503356 [hep-ph]}}.

\bibitem{pythia}
T.~Sjostrand, S.~Mrenna, and P.~Z. Skands,
  \href{http://dx.doi.org/10.1088/1126-6708/2006/05/026}{{\em J. High Energy
  Phys.} {\bfseries 0605} (2006) 026},
\href{http://arxiv.org/abs/hep-ph/0603175}{{\ttfamily arXiv:hep-ph/0603175
  [hep-ph]}}.

\bibitem{Lansberg:2012ta}
J.-P. Lansberg, \href{http://dx.doi.org/10.1016/j.nuclphysa.2012.12.051}{{\em
  Nucl. Phys. A} {\bfseries 470} (2013) 910},
\href{http://arxiv.org/abs/1209.0331}{{\ttfamily arXiv:1209.0331 [hep-ph]}}.

\bibitem{Aaij:2012se}
R.~Aaij {\em et~al.} ({LHCb Collaboration}),
  \href{http://dx.doi.org/10.1007/JHEP11(2012)031}{{\em J. High Energy Phys.}
  {\bfseries 1211} (2012) 031},
\href{http://arxiv.org/abs/1209.0282}{{\ttfamily arXiv:1209.0282 [hep-ex]}}.

\bibitem{Affolder:1999wm}
T.~Affolder {\em et~al.} ({CDF Collaboration}),
  \href{http://dx.doi.org/10.1103/PhysRevLett.84.2094}{{\em Phys. Rev. Lett.}
  {\bfseries 84} (2000) 2094},
\href{http://arxiv.org/abs/hep-ex/9910025}{{\ttfamily arXiv:hep-ex/9910025
  [hep-ex]}}.

\bibitem{Wang:2012is}
K.~Wang, Y.~Q. Ma and K.~T. Chao,
  \href{http://dx.doi.org/10.1103/PhysRevD.85.114003}{{\em Phys. Rev. D}
  {\bfseries 85} (2012) 114003},
\href{http://arxiv.org/abs/1202.6012}{{\ttfamily arXiv:1202.6012 [hep-ph]}}.

\bibitem{Wang:2014_ALICE}
K.~Wang, Y.~Q. Ma and K.~T. Chao, {\em private communication}.

\bibitem{Kisslinger:2013mev}
L.~S. Kisslinger and D.~Das,
  \href{http://dx.doi.org/10.1142/S0217732313501204}{{\em Mod. Phys. Lett. A}
  {\bfseries 28} (2013) 1350120},
\href{http://arxiv.org/abs/1306.6616}{{\ttfamily arXiv:1306.6616 [hep-ph]}}.



\end{thebibliography}
\end{document}